\documentclass{IEEEtran}
\usepackage{amsmath,amsfonts,amssymb}
\usepackage{multicol,multirow,diagbox,bm}
\usepackage{graphicx,subfigure}
\usepackage{lineno,cases}
\allowdisplaybreaks % allow the equation display in different pages
\usepackage{pgf,tikz}
\usetikzlibrary{arrows,shapes,automata,positioning}
\usetikzlibrary{fit}
\usepackage{pst-node}
\usepackage{mathptmx}
\usepackage{cite}
\usepackage{color,xcolor}
\renewcommand{\bar}{\overline}
\renewcommand{\top}{\intercal}

\newcommand{\e}{\epsilon}
\newcommand{\R}{\mathbb{R}}

\newtheorem{assj}{Assumption}
\newtheorem{lemj}{Lemma}
\newtheorem{thmj}{Theorem}
\newtheorem{remj}{Remark}

\newtheorem{defnj}{Definition}
\newtheorem{prob}{Problem}
\newcommand{\pb}{\begin{IEEEproof} }
\newcommand{\pe}{\end{IEEEproof}}

\usepackage{cite}%按照顺序引用
\begin{document}
	\title{Nash Equilibrium Seeking for  High-order Multi-agent Systems with Unknown Dynamics  \thanks{This work was supported in part by National Natural Science Foundation of China under Grants 61973043.}
	}
	\author{Yutao Tang and Peng Yi
		\thanks{Y. Tang is with the School of Artificial Intelligence, Beijing University of Posts and Telecommunications, Beijing 100876, China (e-mail: yttang@bupt.edu.cn). P. Yi is with Department of Control Science and Engineering and also Shanghai Institute of Intelligent Science and Technology, Tongji University, Shanghai, 200092, China (e-mail: yipeng@tongji.edu.cn). }
	}
	\maketitle
	
\begin{abstract}
	In this paper, we consider a Nash equilibrium seeking problem for a class of high-order multi-agent systems with unknown dynamics. Different from existing results for single integrators, we aim to steer the outputs of this class of uncertain high-order agents to the Nash equilibrium of some noncooperative game in a distributed manner. To overcome the difficulties brought by the high-order structure, unknown nonlinearities, and the regulation requirement, we first introduce a virtual player for each agent and solve an auxiliary noncooperative game for them. Then, we develop a distributed adaptive protocol by embedding this auxiliary game dynamics into some proper tracking controller for the original agent to resolve this problem. We also discuss the parameter convergence problem under certain persistence of excitation condition. The efficacy of our algorithms is verified by   numerical examples.   
\end{abstract}

	

\IEEEpeerreviewmaketitle

\section{Introduction}

Nash equilibrium computation is one of the most fundamental problems in noncooperative game theory \cite{basar1999dynamic}. Due to the rapid development of multi-robot networks, machine learning, and big data technologies,  how to develop distributed algorithms to seek a Nash equilibrium has become a hot topic over the past few years. Many important results have been obtained under various circumstances, see \cite{stankovic2011distributed,salehisadaghiani2016distributed, koshal2016distributed,lou2016nash, ye2017distributed, zeng2019generalized, gadjov2019passivity, de2019distributed,yi2019operator} and references therein.

In most continuous-time Nash equilibrium seeking results, the players are assumed to be single integrators. However, in many practical applications, Nash equilibrium seeking might be implemented by or depend upon engineering multi-agent systems with physical dynamics, e.g., \cite{stankovic2011distributed,zhu2013coverage}. Note that these engineering multi-agent systems might not be well modeled as single integrators. Thus it is crucial to consider the Nash equilibrium seeking problem for non-single-integrator multi-agent systems. Although efforts have been made by some authors for full-information circumstances, e.g., \cite{frihauf2011nash,laraki2013higher,fabiani2019nash,ibrahim2019nash},  there are very few works on the solvability of distributed Nash equilibrium seeking problem under the partial information scenario for high-order multi-agent systems.

Recently, some interesting attempts have been made along this line and several classes of non-single-integrator multi-agent systems have been discussed to reach a steady-state related to Nash equilibria of some noncooperative games. In \cite{romano2019dynamic}, the gradient-play rules were extended to solve the Nash equilibrium seeking problem for (multiple) integrators with disturbance rejection. In \cite{bianchi2019continuous}, the authors considered a generalized Nash equilibrium seeking problem with coupling constraints and solved it for double-integrator multi-agent systems.  With regarding to special types of aggregative games, more general agent dynamics have also been explored. For example, passive nonlinear second-order agents were considered in \cite{de2019feedback} by a proportional integral feedback algorithm to reach the expected Cournot-Nash equilibrium. Without assuming the exact knowledge of agent dynamics,  \cite{deng2019distributed} and \cite{zhang2019distributed} further took parameter uncertainties into consideration and developed effective distributed rules to drive the outputs of agents in the Euler-Lagrange form and output feedback form with unity relative degree to reach the Nash equilibrium of some aggregative games.

Motivated by the aforementioned observations,  we consider a noncooperative game played by a class of nonlinear high-order multi-agent systems with unknown dynamics. More specifically, we focus on the case when the unknown time-varying dynamics can be linearly parameterized. As discussed in \cite{romano2019dynamic, deng2019distributed, zhang2019distributed}, we aim to drive the output of this high-order nonlinear multi-agent system to reach a Nash equilibrium specified by the given noncooperative game irrespective of the unknown dynamics. 

The contribution of this paper is at least two-fold. On the one hand, we formulate and solve a Nash equilibrium seeking problem for a class of high-order nonlinear multi-agent systems subject to unknown dynamics. When such unknown dynamics vanishes, the agent dynamics can include both single and multiple integrators as special cases. Thus, this work can be taken as an adaptive high-order extension of the results obtained in \cite{ye2017distributed,gadjov2019passivity,romano2019dynamic}. %,tang2020nash}. 
On the other hand, we develop a novel embedded control design to solve such a Nash equilibrium seeking problem for high-order multi-agent systems, which can substantially reduce the design complexities. By introducing a virtual player for each agent, we convert the original problem into two simpler subproblems, i.e., Nash equilibrium seeking for single-integrator multi-agent systems and output tracking for the considered uncertain high-order multi-agent systems. This treatment facilitates us to solve our Nash equilibrium seeking problem for high-order agents in a modular way.

The rest of this paper is organized as follows. Some preliminaries are provided in Section \ref{sec:pre}. Problem formulation is presented in Section \ref{sec:form}. Then the main results are given in Section  \ref{sec:main} along with both solvability analysis and parameter convergence. Following that, several examples are provided to illustrate the effectiveness of our algorithms in Section \ref{sec:simu}. Finally, concluding remarks are given in Section \ref{sec:con}.
 
\section{Preliminary}\label{sec:pre}
	
In this section, we present some preliminaries of convex analysis and graph theory for the following analysis. More details can be found in   \cite{ruszczynski2006nonlinear} and \cite{godsil2001algebraic}.
	
Let $\R^N$ be the $N$-dimensional Euclidean space and $\R^{N_1\times N_2}$ be the set of all $N_1\times N_2$ matrices. ${\bf 1}_N$ (or ${\bf 0}_N$) denotes  an $N$-dimensional all-one (or all-zero) column vector and ${\bm 1}_{N_1\times N_2}$ (or ${\bm 0}_{N_1\times N_2}$) all-one (or all-zero) matrix. $\mbox{diag}\{b_1,\,{\dots},\,b_N\}$ denotes an $N \times N$ diagonal matrix with diagonal elements $b_i$ with $i=1,\,{\dots},\,N$. $\mbox{blockdiag}(A_1,\,\dots,\,A_N)$ denotes a block diagonal matrix with diagonal elements $A_i$ with $i=1,\,\dots,\,N$. $\mbox{col}(a_1,\,{\dots},\,a_N) = [a_1^\top,\,{\dots},\,a_N^\top]^\top$ for column vectors $a_i$ with $i=1,\,{\dots},\,N$.  For a vector $x$ and a matrix $A$, $||x||$ denotes the Euclidean norm and $||A||$ the spectral norm. Let $M_1=\frac{1}{\sqrt{N}}{\bm 1}_N$ and $M_2\in \R^{N\times (N-1)}$ be the matrix satisfying $M_2^\top M_1={\bm 0}_{N-1}$, $M_2^\top M_2=I_{N-1}$ and $M_2 M_2^\top=I_{N}-M_1 M_1^\top$.  We may omit the subscript when it is self-evident.  % For a square matrix $A$.  $\mbox{Tr}(A)$ denotes the trace of $A$.% and $||A||_{\rm F}=\mbox{Tr}(A^\top A)$ denotes its Frobenius norm.
	
A function $f\colon \R^m \rightarrow \R $ is said to be convex if, for any $0\leq a \leq 1$ and $\zeta_1,\zeta_2 \in \R^m$, we have $f(a\zeta_1+(1-a)\zeta_2)\leq af(\zeta_1)+(1-a)f(\zeta_2)$. It is said to be strictly convex if this inequality is strict whenever $\zeta_1 \neq \zeta_2$. A vector-valued function $\Phi \colon \R^m \rightarrow \R^m$ is said to be  $\omega$-strongly monotone, if for any $\zeta_1,\, \zeta_2 \in \R^m$,  we have $(\zeta_1-\zeta_2)^\top [\Phi(\zeta_1)-\Phi(\zeta_2)]\geq \omega \|\zeta_1-\zeta_2\|^2$.  Function $\Phi\colon \R^m \rightarrow \R^m$ is said to be  $\vartheta$-Lipschitz, if for any $ \zeta_1, \, \zeta_2 \in \R^m$, it holds that $\|\Phi(\zeta_1)-\Phi(\zeta_2)\|\leq \vartheta \|\zeta_1-\zeta_2\|$.

A weighted directed graph (or digraph) $\mathcal {G}=(\mathcal {N}, \mathcal {E}, \mathcal{A})$ is defined as follows, where $\mathcal{N}=\{1,\,{\dots},\,N\}$ is the set of nodes, $\mathcal {E}\subset \mathcal{N}\times \mathcal{N}$ is the set of edges, and $\mathcal{A}\in \mathbb{R}^{N\times N}$ is a weighted adjacency matrix. $(i,j)\in \mathcal{E}$ denotes an edge leaving from node $i$ and entering node $j$. The weighted adjacency matrix of this digraph $\mathcal {G}$ is described by $A=[a_{ij}]\in \R^{n\times n}$, where $a_{ii}=0$ and $a_{ij}\geq 0$ ($a_{ij}>0$ if and only if there is an edge from agent $j$ to agent $i$).  A path in graph $\mathcal {G}$ is an alternating sequence $i_{1}e_{1}i_{2}e_{2}{\cdots}e_{k-1}i_{k}$ of nodes $i_{l}$ and edges $e_{m}=(i_{m},\,i_{m+1})\in\mathcal {E}$ for $l=1,\,2,\,{\dots},\,k$. If there exists a path from node $i$ to node $j$ then node $i$ is said to be reachable from node $j$. The neighbor set of agent $i$ is defined as $\mathcal{N}_i=\{j\colon (j,\,i)\in \mathcal {E} \}$ for $i\in \mathcal{N}$.  A graph is said to be undirected if $a_{ij}=a_{ji}$ ($i,\,j=1,\,{\dots},\,N$).  If there is a directed path between any two nodes, then the digraph is said to be strongly connected.   An undirected graph is said to be connected for short if it is strongly connected.

The in-degree and out-degree of node $i$ are defined by $d^{\mbox{in}}_i=\sum\nolimits_{j=1}^N a_{ij}$ and $d^{\mbox{out}}_i=\sum\nolimits_{j=1}^N a_{ji}$. A digraph is weight-balanced if $d^{\mbox{in}}_i=d^{\mbox{out}}_i$ holds for any $i=1,\,\dots,\,N$.  The Laplacian matrix of $\mathcal{G}$ is defined as $L\triangleq D^{\mbox{in}}-\mathcal{A}$ with $D^{\mbox{in}}=\mbox{diag}(d^{\mbox{in}}_1,\,\dots,\,d^{\mbox{in}}_N)$.  Note that $L{\bm 1}_N={\bm 0}_N$ for any digraph. When it is weight-balanced, we have ${\bm 1}_N^\top L={\bm 0}_N^\top$ and the matrix  $\mbox{Sym}(L)\triangleq \frac{L+L^\top}{2}$ is positive semidefinite. Then, we can order the eigenvalues of $\mbox{Sym}(L)$ as $0=\lambda_1<\lambda_2\leq \dots\leq \lambda_N$.  A weight-balanced digraph is strongly connected if and only if $\lambda_2>0$. In this case, we have $\lambda_2 I_{N-1}\leq M_2^\top \mbox{Sym}(L)M_2\leq \lambda_N I_{N}$. %See \cite{godsil2001algebraic} for more details. 

\section{Problem formulation} \label{sec:form}

In this paper, we consider a collection of heterogeneous high-order nonlinear systems described by: 
	\begin{align}\label{sys:agent}
	\begin{split}
	\dot{x}_{j,i}&=x_{j+1,i}\\ 
	\dot{x}_{n_i,i}&=\theta_i^\top {\bm p}_{i}(x_{i},\,t)+u_i\\ 
	y_i&=x_{1,i},\quad i=1,\,\dots,\,N,\,j=1,\,\dots,\,n_i-1
	\end{split}
	\end{align}
where $x_{j,i} \in \R $ is the $j$-th state variable of agent $i$, $x_i\triangleq \mbox{col}(x_{1,i},\,\dots,\,x_{n_i,i}) \in \R^{n_i}$, $y_i \in \mathbb{R}$ and $u_i \in \mathbb{R}$ are respectively the state, output, and input of agent $i$. The function $\Delta_i(x_i,\, \theta_i,\,t) \triangleq \theta_i^\top {\bm p}_{i}(x_{i},\,t)$ represents the unknown time-varying nonlinearity which might result from modeling errors or external perturbations with uncertain parameter vector $\theta_i=\mbox{col}(\theta_{1,i},\, \dots,\,\theta_{n_{\theta_i},i})\in \R^{n_{\theta_i}}$ and known basis function vector ${\bm p}_i(x_i,\,t)$. Furthermore, we assume that the basis function ${\bm p}_i(x_i,\,t)$ can be uniformly bounded by smooth functions of $x_i$, which trivially holds when it is time-invariant.

Equation \eqref{sys:agent} can represent a plenty of practical systems and  is general enough to cover integrators, Van der Pol systems, Duffing equations and many mechanical systems \cite{khalil2002nonlinear}. Moreover, this time-varying feature can  be further utilized to model many typical external disturbances, e.g.  constants and sinusoidal signals.  Note that when $n_{\theta_i}=0$, the unknown nonlinearity vanishes and this class of agent dynamics include both single and multiple integrators as special cases.

Suppose these agents play a $N$-player noncooperative game defined as follows. Agent $i$ is endowed with a continuously differentiable cost function ${J}_i(y_i,\,y_{-i})$, where $y_i\in \R$ denotes the  output  strategy profile of agent $i$ specified by \eqref{sys:agent} and  $y_{-i}\in \R^{N-1}$ denote the output strategy profile of this multi-agent system except for agent $i$. In this game, each player seeks to asymptotically minimize its own cost function $J_i$ by reaching some proper steady-state output strategy.

The equilibrium point of this game is defined as in \cite{basar1999dynamic,gadjov2019passivity}.
\begin{defnj}\label{def:prop}
	%\em
	Consider the game $\mbox{G}=\{\mathcal{N},\,{J}_i,\,\R\}$.  A strategy profile $y^*=\mbox{col}(y_1^*,\,\dots,\,y_N^*)$ is said to be a Nash equilibrium of $\mbox{G}$ if	$J_i(y_i^*,\,y_{-i}^*)\leq J_i(y_i,\,y_{-i}^*)$ for any $i\in \mathcal{N}$ and $y_i\in \R$.
\end{defnj}
	
At a Nash equilibrium of the game $\mbox{G}$, all agents tend to keep at this state since no player can unilaterally decrease its cost by changing the steady-state output strategy on its own.  Denote $\nabla_i J_i(y_i,\,y_{-i})\triangleq \frac{\partial }{\partial y_i}J_i(y_i,\,y_{-i})\in \R$ and $F(y)\triangleq \mbox{col}(\nabla_1 J_1(y_1,y_{-1}),\dots,\nabla_N J_N(y_N,y_{-N}))\in \R^N$.  Here, $F$ is called the pseudogradient associated with $J_1,\,\dots,\,J_N$.  

The following assumption is often made in Nash equilibrium seeking literature \cite{gadjov2019passivity,deng2019distributed,de2019distributed}.

\begin{assj}\label{ass:convex+monotone}
	%\em
	For any $i \in \mathcal{N}\triangleq \{1,\,\dots,\,N\}$,  function $J_i(y_i,\,y_{-i})$ is twice continuously differentiable, strictly convex, and radially unbounded in $y_i\in \R$ for any fixed $y_{-i}\in \R^{N-1}$. Moreover, the associated pseudogradient $F$ is assumed to be $\underline l$-strongly monotone and $\bar {l}$-Lipschitz for some constants $\underline{l},\, \bar l>0$. 	
\end{assj}
 
Under this assumption, this noncooperative game $\mbox{G}$ admits a unique Nash equilibrium $y^*$ which can be characterized by the equation $F(y^*)={\bm 0}$ according to Propositions 1.4.2 and 2.2.7 in \cite{facchinei2003finite}. To seeking this Nash equilibrium $y^*$, we are interested in distributed designs where each agent only gets the information of a subset of the overall agents. For this purpose, a digraph $\mathcal{G}=(\mathcal{N},\, \mathcal{E}, \,\mathcal{A})$  is used to describe the information flow among these agents with a node set $\mathcal{N}=\{1,\,\dots,\, N\}$ and a weighted matrix $\mathcal{A}=[a_{ij}]_{N\times N}$. An edge $(i,\, j)$ in digraph $\mathcal{G}$ means that agent $j$ can the information of agent $j$.  

The following assumption is often made in multi-agent coordination literature \cite{ren2008distributed,ye2017distributed,gadjov2019passivity}.
\begin{assj}\label{ass:graph}	%\em
	Digraph $\mathcal{G}$ is weight-balanced and strongly connected. 
\end{assj}

The distributed Nash equilibrium seeking problem considered in this paper is readily formulated as follows.
\begin{prob}
	%\em
	For given multi-agent system  \eqref{sys:agent}, digraph $\mathcal{G}$, and function $J_i$, determine a distributed protocol $u_i$ for agent $i$ using its own local data and exchanged information with its neighbors such that
	\begin{itemize}
		\item[1)] all trajectories of the closed-loop system are bounded over the time interval $[0,\,+\infty)$;
		\item[2)] the outputs of agents satisfy $\lim_{t\to +\infty}||y_i(t)-y_i^*||=0$ for any $i\in \mathcal{N}$ with $y^*=\mbox{col}(y_1^*,\,\dots,\,y^*_N)$ being the Nash equilibrium of  game $\mbox{G}$.
	\end{itemize}
\end{prob}

\begin{remj}
	The formulated problem has been studied by many authors when the agent dynamics is restricted to single and/or multiple integrators. In contrast with existing works, the considered agents in this paper are allowed to be high-order heterogeneous subject to unknown dynamics. These features make our problem much more challenging than existing Nash equilibrium seeking works for integrators. 
\end{remj}

In fact, it can be further found that the main difficulty to solve the Nash equilibrium seeking problem for these agents lies in the couplings among the high-order structure, unknown nonlinearities, and the global equilibrium regulation requirement.  	Inspired by the embedded control scheme  developed in \cite{tang2019cyb,tang2020optimal}  to solve distributed optimization problems, we borrow this decoupling idea to reduce these design complexities and extend it to solve the formulated  Nash equilibrium seeking problem for high-order multi-agent systems \eqref{sys:agent} by developing novel adaptive controllers in the following section.
	
\section{Main Result }\label{sec:main}

In this section, we will detail the main design to solve our Nash equilibrium seeking problem along with parameter convergence analysis.

\subsection{Embedded  design}

Motivated by the embedded control scheme in \cite{tang2019cyb} and \cite{tang2020optimal}, we first consider some virtual multi-agent system 
\begin{align}
\dot{z}_i=\mu_i,\quad i\in \mathcal{N}
\end{align}
Suppose each virtual agent $i$ is assigned with the same cost function $J_i$ as agent $i$ and plays the same noncooperative game as in Problem \eqref{def:prop} in order to asymptotically reproduce the Nash equilibrium $y^*$ by the vector $\mbox{col}(z_1,\,\dots,\,z_N)$. Here, each virtual player can be understood as an abstraction of the original agent \eqref{sys:agent} as discussed in \cite{girard2009hierarchical,tang2013hierarchical}. Then, agent \eqref{sys:agent}  takes $z_i$ as its output reference to reach the expected Nash equilibrium point. In this way, the Nash equilibrium seeking problem for agents with complex dynamics is divided into two simpler subproblems, i.e., Nash equilibrium seeking problem for single integrators and reference tracking problem for original agent, which can be independently solved in a modular way.

Note that the first subproblem is essentially a conventional Nash equilibrium seeking problem  and has been well-studied in existing literature \cite{gadjov2019passivity,yi2019operator,ye2017distributed,de2019distributed}. We use the following dynamics for the virtual agent $i$, which is a distributed version of gradient-play rules for game $\mbox{G}$:
\begin{align}\label{sys:generator}
\begin{split}
\dot{z}_i&=-\alpha \sum\nolimits_{j=1}^N a_{ij}({z}_i-{{z}}^j_i)-\nabla_i J_i({\bf z}^i)\\
\dot{z}_k^i&=-\alpha \sum\nolimits_{k=1}^N a_{ij}({{z}}_k^i-{{z}}_k^j),\quad k \in \mathcal{N}\backslash \{i\}
\end{split}
\end{align}
where ${{\bf z}}^i=\mbox{col}(z_1^i,\,\dots,\,z_N^{i})$ represents agent $i$'s estimate of all virtual agents' strategies with $z_i^i=z_i$ and the constant $\alpha>0$ is to be specified later. The function $\nabla_i J_i({\bf z}^i)=\frac{\partial J_i}{\partial z_i^i}({\bf z}^i)$ is the partial gradient of agent $i$'s cost function evaluated at the local estimate ${\bf z}^i$. 

For convenience, we define an extended pseudogradient as ${\bm F}({\bf z})=\mbox{col}( \nabla_1 J_1({\bf z}^1),\,\dots,\, \nabla_N J_N({\bf z}^N))\in \R^N$.  The following assumption has been made in \cite{gadjov2019passivity, de2019distributed}.
\begin{assj}\label{ass:lip-extended}
	The extended pseudogradient $\bm F$ is $l_{F}$-Lipschitz with $l_{F}>0$.
\end{assj}

Putting \eqref{sys:generator} into a compact form gives
\begin{align}\label{sys:generator-compact}
\dot{{\bf z}}=-\alpha {\bf L}{\bf z}-R {\bm F}({\bf z})
\end{align}
where ${\bf z}=\mbox{col}({\bf z}^1,\,\dots,\,{\bf z}^N)$, $R=\mbox{diag}(R_1,\,\dots,\,R_N)$, $R_i=\mbox{col}({\bm 0}_{i-1},\,1,\,{\bm 0}_{N-i})$, and ${\bf L}=L\otimes I_N$ with the extended pseudogradient ${\bm F}({\bf z})$.  

Denote $l=\max\{\bar l,\, l_{F}\}$. When $\alpha=1$, system \eqref{sys:generator-compact} reduces to the consensus-based gradient-play dynamics in \cite{gadjov2019passivity}. Here we add an adjustable parameter $\alpha$ to increase the gain of the proportional term ${\bf L} {\bf z}$. With this gain being large enough, the effectiveness of algorithm \eqref{sys:generator-compact} can be established as follows.
\begin{lemj}\label{lem:generator}
		%\em
	Suppose Assumptions \ref{ass:convex+monotone}--\ref{ass:lip-extended} hold. Let 
%	\begin{align}\label{eq:parameter-alpha}
$\alpha>\frac{1}{\lambda_2}(\frac{l^2}{\underline{l}}+l)$. 
%	\end{align}
	Then, for any $i\in \mathcal{N}$, along the trajectory of system \eqref{sys:generator-compact}, ${\bm z}^i(t)$ exponentially converges to $y^*$ as $t$ goes to $+\infty$.
\end{lemj}

\pb  We first show that at the equilibrium of system \eqref{sys:generator-compact}, $z_i$ indeed reaches the Nash equilibrium of game $\mbox{G}$. In fact, letting the righthand side of \eqref{sys:generator} be zero, we have $\alpha {\bf L}{\bf z}^*+ R {\bm F}({\bf z}^*)={\bm 0}$. Premultiplying  both sides by ${\bm 1}^\top_N \otimes I_N$ gives
\begin{align*}
{\bm 0}=\alpha ({\bm 1}^\top_N \otimes I_N)(L\otimes I_N){\bf z}^*+ ({\bm 1}^\top_N \otimes I_N) R {\bm F}({\bf z}^*)
\end{align*}
Using ${\bm 1}^\top_NL=0$ gives 
%\begin{align*}
${\bm 0}=({\bm 1}^\top_N \otimes I_N) R {\bm F}({\bf z}^*)$. 
%\end{align*}
By the notation of $R$ and $\bm F$, we have ${\bm F}({\bf z}^*)=\bm 0$. This further implies that ${\bf L}{\bf z}^*=\bm 0$. Recalling the property of $L$ under Assumption \ref{ass:graph}, one can determine some $\theta \in \R^N$ such that ${\bf z}^*={\bm 1}\otimes \theta $. This means ${\bm F}({\bm 1}\otimes \theta )=\bm 0$ and thus $\nabla_i J_i(\theta_i,\,\theta_{-i})=0$, or equivalently, $F(\theta)=\bm 0$. That is, $\theta$ is the unique Nash equilibrium $z^*$ of $\mbox{G}$ and ${\bf z}^*={\bm 1}\otimes z^* $.

Next, we show the exponential stability of system \eqref{sys:generator-compact} at its equilibrium ${\bf z}^*={\bm 1}\otimes z^*$. For this purpose, we denote $\tilde  {\bf z}={\bf z}-{\bf z}^*$ and perform the coordinate transformation $\bar {\bf z}_1=(M_1^\top\otimes I_N)\tilde  {\bf z}$ and $\bar {\bf z}_2=(M_2^\top\otimes I_N)\tilde  {\bf z}$. It follows that
\begin{align*}
\dot{\bar{\bf z}}_1&=-(M_1^\top\otimes I_N)R \Delta\\
\dot{\bar {\bf z}}_2&=-\alpha [(M_2^\top L M_2)\otimes I_{N}]\bar {\bf z}_2-(M_2^\top\otimes I_N) R\Delta
\end{align*} 
where  $\Delta\triangleq {\bm F}({\bf z})- {\bm F}({\bf z}^*)$. 

Let $V_0(\bar {\bf z}_1,\,\bar {\bf z}_2)=\frac{1}{2}(||\bar {\bf z}_1||^2+||\bar {\bf z}_2||^2)$. Then, its time derivative along the trajectory of system \eqref{sys:generator-compact} satisfies that
\begin{align}\label{eq1:lem-generator}
\dot{V}_0&=-\bar {\bf z}_1^\top (M_1^\top\otimes I_N)R \Delta-\bar {\bf z}_2^\top (M_2^\top\otimes I_N) R\Delta\nonumber\\
&- \alpha \bar {\bf z}_2^\top  \{[M_2^\top  L  M_2]\otimes I_{N}\}\bar {\bf z}_2\nonumber\\
&=-\tilde {\bf z}^\top  R \Delta - \alpha \bar {\bf z}_2^\top  [(M_2^\top \mbox{Sym}(L) M_2)\otimes I_{N}]\bar {\bf z}_2\nonumber\\
&\leq -\alpha \lambda_2 ||\bar {\bf z}_2||^2-\tilde {\bf z}^\top R \Delta
\end{align}
Since $\tilde {\bf z}=(M_1 \otimes I_N) \bar {\bf z}_1+(M_2 \otimes I_N) \bar {\bf z}_2\triangleq \tilde {\bf z}_1+\tilde {\bf z}_2$, we split $\tilde {\bf z}$ into two parts to estimate the above cross term and obtain that
\begin{align*}
-\tilde {\bf z}^\top R \Delta&=(\tilde {\bf z}_1+\tilde {\bf z}_2 )^\top R [{\bm F}(\tilde {\bf z}_1+\tilde {\bf z}_2+{\bf z}^*)- {\bm F}({\bf z}^*)]\\
&=- \tilde {\bf z}_1^\top R [{\bm F}(\tilde {\bf z}_1+\tilde {\bf z}_2+{\bf z}^*)-{\bm F}(\tilde {\bf z}_1+{\bf z}^*)]\\
&-\tilde {\bf z}_2^\top R [{\bm F}(\tilde {\bf z}_1+\tilde {\bf z}_2+{\bf z}^*)-{\bm F}(\tilde {\bf z}_1+{\bf z}^*)]\\
&-\tilde {\bf z}_1^\top R [{\bm F}(\tilde {\bf z}_1+{\bf z}^*)-{\bm F}({\bf z}^*)]\\  %负定项
&-\tilde {\bf z}_2^\top R [{\bm F}(\tilde {\bf z}_1+{\bf z}^*)-{\bm F}({\bf z}^*)]
\end{align*}
As we have ${\bm F}({\bm 1}_N\otimes y)=F(y)$ for any $y\in \R^N$, it follows by the strong monotonicity of $F$  that
\begin{align*}
&\tilde {\bf z}_1^\top R [{\bm F}(\tilde {\bf z}_1+{\bf z}^*)-{\bm F}({\bf z}^*)]\\
&\quad =\frac{\bar {\bf z}_1^\top}{\sqrt{N}} [{\bf F}({\bf 1}\otimes (\frac{\bar {\bf z}_1}{\sqrt{N}}+y^*))-{\bm F}({\bf 1}\otimes y^*)]\\
&\quad =\frac{\bar {\bf z}_1^\top}{\sqrt{N}} [F(y^*+\frac{{\bar {\bf z}}_1}{\sqrt{N}})-{F}(y^*)]\\
%&\quad =\frac{\bar {\bf z}_1^\top}{\sqrt{N}} [F(y^*+\frac{{\bar {\bf z}}_1}{\sqrt{N}})-{F}(y^*)]\\
&\quad \geq \frac{\underline{l}}{N}||{\bar {\bf z}}_1||^2
\end{align*} 
where we use the identity $({\bf 1}^\top \otimes I_N) R=I_N$ and $\tilde {\bf z}_1^\top  R =\frac{\bar {\bf z}_1^\top }{\sqrt{N}}$. Note that $||R||=||M_2||=1$ by definition. This implies that $||R^\top \tilde {\bf z}_2 ||\leq  ||\tilde {\bf z}_2|| =||\bar {\bf z}_2||$. Then, under Assumptions \ref{ass:convex+monotone} and \ref{ass:lip-extended}, we have that
%\begin{align}\label{eq2:lem-generator}
%-\tilde {\bf z}^\top R \Delta&\leq \frac{2 l}{\sqrt{N}}||\bar {\bf z}_1||||\bar {\bf z}_2||+l||\bar {\bf z}_2||^2-\frac{\underline{l}}{N}||\bar {\bf z}_1||^2 \nonumber \\
%&\leq -\frac{\underline{l}}{2N}||\bar {\bf z}_1||^2+(\frac{2 {l}^2}{\underline{l}}+l)||\bar {\bf z}_2||^2 \nonumber \\
%&\leq -\frac{\underline{l}}{2N}||\bar {\bf z}_1||^2+\frac{3 {l}^2}{\underline{l}} ||\bar {\bf z}_2||^2
%\end{align}
\begin{align}\label{eq2:lem-generator}
-\tilde {\bf z}^\top R \Delta&\leq \frac{2 l}{\sqrt{N}}||\bar {\bf z}_1||||\bar {\bf z}_2||+l||\bar {\bf z}_2||^2-\frac{\underline{l}}{N}||\bar {\bf z}_1||^2 \end{align}
Bringing inequalities \eqref{eq1:lem-generator} and \eqref{eq2:lem-generator} together gives
\begin{align}\label{eq3:lem-generator}
\dot{V}_0&\leq -\frac{\underline{l}}{N}||\bar {\bf z}_1||^2 -(\alpha \lambda_2- l)||\bar {\bf z}_2||^2+\frac{2 l}{\sqrt{N}}||\bar {\bf z}_1||||\bar {\bf z}_2|| \nonumber \\ 
&=-\begin{bmatrix}
||\bar {\bf z}_1||& 
||\bar {\bf z}_2||
\end{bmatrix} 
A_\alpha \begin{bmatrix}
||\bar {\bf z}_1||\\
||\bar {\bf z}_2||
\end{bmatrix}
\end{align}
with a matrix $A_\alpha=\begin{bmatrix} \frac{\underline l}{N}&-\frac{l}{\sqrt{N}}\\ -\frac{l}{\sqrt{N}}&\alpha \lambda_2-l \end{bmatrix}$. Note that when $\alpha>\frac{1}{\lambda_2}(\frac{l^2}{\underline{l}}+l)$, this matrix $A_\alpha$ is positive definite. Thus, there must be a constant $\nu>0$ such that
\begin{align*}%\label{eq3:lem-generator}
\dot{V}_0&\leq -\nu V_0
\end{align*}
Using Theorem 4.10 in \cite{khalil2002nonlinear}, one can conclude the exponential convergence of ${\bf z}(t)$ to ${\bf z}^*$, which implies that ${\bf z}^i(t)$ converges to $z^*$ as $t\to +\infty$. The proof is thus complete.
\pe

\begin{remj}
 The criterion to choose $\alpha$ clearly presents a natural trade-off between the control efforts and graph algebraic connectivity. This observation is consistent with the results in \cite{gadjov2019passivity} when $\alpha$ is fixed as one. By choosing a large enough $\alpha$, this lemma ensures the exponential convergence of virtual agent's state to the Nash equilibrium $z^*$ under weight-balanced digraphs and also provides an alternative way to remove the restrictive graph coupling condition other than singular perturbation analysis in \cite{gadjov2019passivity}.  
\end{remj}

With this lemma, we are left to solve the output tracking problem for agent \eqref{sys:agent} with reference $z_i(t)$. 

Due to the presence of uncertain parameter $\theta_i$, direct cancellation technique cannot be used to handle the nonlinearities  in \eqref{sys:agent}.  To tackle this issue, we adopt a certainty-equivalence design and propose an adaptive controller for each agent:
	\begin{align}\label{ctr:offline}
	u_{i}&=-\hat \theta_i^\top {\bm p}_{i}(x_{i},\,t)+\frac{1}{\e^{n_i}}[k_{1i}(x_{1,i}-z_i)+\sum_{j=2}^{n_i}\e^{j-1}k_{ji}x_{j,\,i}] \nonumber\\
	\dot{\hat \theta}_i&=\phi_i(x_i,\,\hat \theta_i,\, z_i,\,t) \nonumber\\
	\dot{z}_i&=-\alpha \sum_{j=1}^N a_{ij}({z}_i-{{z}}^j_i)-\nabla_i J_i({\bf z}^i)\\
	\dot{z}_k^i&=-\alpha \sum_{k=1}^N a_{ij}({{z}}_k^i-{{z}}_k^j),\quad k \in \mathcal{N}\backslash  \{i\} \nonumber
	\end{align} 
where $\hat \theta_i $ is the estimation of $\theta_i$ with constants $\e$,\,$k_{1i},\,\dots,\,k_{n_i i}>0$ and smooth function $\phi_i(\cdot)$ to be specified later. 
This controller is  distributed in the sense of using each agent's own local data and exchanged information with the neighbors.

Under the above control law, we have:
\begin{align}\label{sys:closed-spform-0}
\begin{split}
\dot{x}_{1,\,i}&=x_{2, i} \\ 
\vdots&\\
\dot{x}_{n_i,i}&=(\theta_i^\top-\hat \theta_i^\top) {\bm p}_{i}(x_{i},\,t)+\frac{1}{\e^{n_i}}[k_{1i}(x_{1,i}-z_i)\\
& +\sum\nolimits_{j=2}^{n_i}\e^{j-1}k_{ji}x_{j,\,i}]\\ 
\dot{\hat \theta}_i&=\phi_i(x_i,\,\hat \theta_i,\,z_i,\,t)\\
	\dot{z}_i&=-\alpha \sum\nolimits_{j=1}^N a_{ij}({z}_i-{{z}}^j_i)-\nabla_i J_i({\bf z}^i)\\
\dot{z}_k^i&=-\alpha \sum\nolimits_{k=1}^N a_{ij}({{z}}_k^i-{{z}}_k^j),\quad k \in \mathcal{N}\backslash  \{i\}
\end{split}
\end{align}

Letting $\hat x_i=\mbox{col}(x_{1,i}-z_i,\,\e x_{2,i},\,\dots,\,\e^{n_i-1}x_{n_i,i})$, we can further rewrite \eqref{sys:closed-spform-0} as follows.
\begin{align}\label{sys:closed-spform}
\begin{split}
\e \dot{\hat x}_i&=A_i\hat x_i-\e b_{1i} \dot{z}_i+{\e^{n_i}}b_{2i}(\theta_i^\top-\hat \theta_i^\top) {\bm p}_{i}(x_{i},\,t) \\
\dot{\hat {\theta}}_i&=\phi_i(x_i,\,\hat \theta_i,\,z_i,\,t)\\
\dot{z}_i&=-\alpha \sum\nolimits_{j=1}^N a_{ij}({z}_i-{{z}}^j_i)-\nabla_i J_i({\bf z}^i)\\
\dot{z}_k^i&=-\alpha \sum\nolimits_{k=1}^N a_{ij}({{z}}_k^i-{{z}}_k^j),\quad k \in \mathcal{N}\backslash  \{i\}
\end{split}
\end{align}
where  $A_i=\left[\begin{array}{c|c}
{\bf 0}&I_{n_{x_i}-1}\\\hline
k_{1i}&[k_{2i}\,\dots\,k_{n_i\,i}]
\end{array}\right]$, $b_{1i}=\mbox{col}(1,\,{\bf 0})$, and $b_{2i}=\mbox{col}({\bf 0},\,1)$.

%Note that the above system in cascading form. 
We first choose constants $k_{1i},\,\dots,\,k_{n_i\,i}$ such that the polynomial $s^{n_i}-k_{n_i\, i}s^{n_i-1}-\dots-k_{2i}s-k_{1i}$ is Hurwitz for any $1\leq i\leq N$. Then, the Lyapunov equation $A_i^\top P_i+P_iA_i=-2I_{n_i}$ has a unique positive definite solution $P_i$ with compatible dimensions for any $i\in \mathcal{N}$. Based on the above observations, we only need to determine some proper function $\phi_i(\cdot)$  such that all trajectories of \eqref{sys:closed-spform} is bounded over $[0,\,+\infty)$ and satisfying $\hat x_{i}(t)\to 0$ as $t$ goes to infinity.

%Note that the above system is almost in a singularly perturbed form  except the adaption dynamics $\hat \theta_i$ \cite{khalil2002nonlinear}. By letting $\e=0$, we have $\hat x_i=0$ and $x_{1,i}=z_i$. The resultant quasi-state-state model of the above composite system is exactly system \eqref{sys:generator}, which in turn guarantees $y_i(t)\to y_i^*$ as $t$ goes to infinity by Lemma 1. 

\subsection{Solvability Analysis}

Denote $\hat x=\mbox{col}(\hat x_1,\,\dots,\,\hat x_N)$,  $\theta=\mbox{col}(\theta_1,\,\dots,\,\theta_N)$,  $\hat \theta=\mbox{col}(\hat \theta_1,\,\dots,\,\hat \theta_N)$, $\bar \theta=\theta-\hat \theta$, and $z=\mbox{col}(z_1,\,\dots,\,z_N)$ for short. The whole composite multi-agent system can be put into a compact form as follows.
	\begin{align}\label{sys:composite-compact}
	\begin{split}
	\dot{\hat x}&=\frac{1}{\e}A\hat  x-B_{1} \dot{z}+EB_{2} {\bm p}^\top (x,\,t) \bar \theta\\
	\dot{\bar {\theta}}&=\phi(x,\,\hat \theta,\,z,\,t) \\
	\dot{{\bf z}}&=-\alpha {\bf L}{\bf z}-R {\bm F}({\bf z})
	\end{split}
	\end{align}
where 
\begin{align*}
&A\triangleq \mbox{blockdiag}(A_1,\,\dots,\,A_N)\\
&B_1\triangleq \mbox{blockdiag}(b_{11},\,\dots,\,b_{1N})\\
&B_2\triangleq \mbox{blockdiag}(b_{21},\,\dots,\,b_{2N})\\
&E\triangleq \mbox{blockdiag}(\e^{n_1-1}I_{n_1},\,\dots,\,\e^{n_N-1}I_{n_N})\\
&\phi(x,\,\hat \theta,\,z,\,t)\triangleq \mbox{col}(\phi_1(x_1,\,\hat \theta_1,\,z_1,\,t), \,\dots,\,\phi_N(x_N,\,\hat \theta_N,\,z_N,\,t))\\
&{\bm p}(x,\,t)\triangleq \mbox{blockdiag}({\bm p}_1(x_1,\,t),\,\dots,\,{\bm p}_N(x_N,\,t))
\end{align*}
	
Here is the first main theorem of this paper.	
\begin{thmj}\label{thm:main}
	Suppose Assumptions \ref{ass:convex+monotone}--\ref{ass:lip-extended} hold. Choose constants $k_{1i},\,\dots,\,k_{n_i\,i}$ such that the polynomial $s^{n_i}-k_{n_i\, i}s^{n_i-1}-\dots-k_{2i}s-k_{1i}$ is Hurwitz for each $ i\in\mathcal{N}$ and let $\alpha>\frac{1}{\lambda_2}(\frac{l^2}{\underline{l}}+l)$. Then, the Nash equilibrium seeking problem for multi-agent system \eqref{sys:agent} is solved by distributed controllers of the form \eqref{ctr:offline} with $\phi_i(x_i,\,\hat \theta_i,\,z_i,\,t)={\bm p}_{i}(x_{i},\,t) b_{2i}^\top P_i \hat x_i$ for any $\e>0$.%, where the constant $\alpha$ is chosen to satisfy \eqref{eq:parameter-alpha}.% and $k_{\iota i }$ are chosen as above for $i=1,\,\dots,\,N,\,\iota =1,\,\dots,\,n_i$.
\end{thmj}
\pb  Under the theorem conditions, we can recall Lemma \ref{lem:generator} and conclude the exponential convergence of $z_i(t)$ towards to $y_i^*$. Then, it is sufficient for us to prove $\hat x_{i}(t)\to 0$ as $t\to +\infty$ to ensure that  $\lim_{t\to+\infty}[y_i(t)-z_i(t)]=0$. To this end, we present a Lyapunov analysis for system \eqref{sys:composite-compact}.

Let us consider the first two subsystems of \eqref{sys:composite-compact}. Let $\hat V_i=\hat W_i+{\e^{n_i-1}}\bar \theta_i^\top \bar \theta_i$ with $\hat W_i=\hat x_i^\top P_i\hat x_i$. Its time derivative along the trajectory of the composite system \eqref{sys:composite-compact} satisfies
\begin{align*}
	\dot{\hat V}_i= &2\hat x_i^\top P_i [\frac{1}{\e}A_i\hat x_i-b_{1i} \dot{z}_i+{\e^{n_i-1}}b_{2i}\bar \theta_i^\top {\bm p}_{i}(x_{i},\,t)]\\
	&-2\e^{n_i-1}\bar \theta_i^\top\phi_i(x_i,\,\hat \theta_i,\,z_i,\,t)\\
	= & -\frac{2}{\e}\hat x_i^\top\hat x_i-2\hat x_i^\top P_i b_{1i} \dot{z}_i 
\end{align*}
	
By Young's inequality, it holds that 
\begin{align*}
	\dot{\hat V}_i \leq &-\frac{2}{\e}\hat x_i^\top\hat x_i+\frac{1}{\e}||\hat x_i||^2+\e||P_i b_{1i}||^2 ||\dot{z}_i||^2 \\
	\leq &-\frac{1}{\e}||\hat x_i||^2+c_1||\dot{z}_i||^2
\end{align*}
where $c_1=\max_{i\in \mathcal{N}}{\e||P_i b_{1i}||^2}$. 

At the same time, we can determine a quadratic Lyapunov function $V_0({\bar {\bf z}})$ according to Lemma \ref{lem:generator} or its proof such that 
$$\dot{V}_0(t)\leq -\nu V_0$$
for some constant $\nu>0$ with $\bar {\bf z }={\bf z }-{\bf 1}\otimes y^*$.  Under Assumption \ref{ass:lip-extended}, the righthand side of system \eqref{sys:generator-compact} is globally Lipschitz. Thus, there exists a constant  $c_2>0$ such that $||\dot{z}_i||^2\leq ||\dot{\bf z}||^2\leq c_2 V_0(\bar {\bf z})$ along the trajectory of system \eqref{sys:generator-compact}.

Next, we choose a Lyapunov function for the whole composite system \eqref{sys:composite-compact} perhaps after some coordinate transformation of ${\bf z}$ to $\bar {\bf z}$ as $\hat V=\sum_{i=1}^N \hat V_i+c_3V_0$ with $c_3>0$ to be specified later. It is positive definite and radially unbounded. Combining the above inequalities, one has
\begin{align*}%\label{eq:thm1-eq1}
	\dot{\hat V} \leq  &-\sum_{i=1}^N \frac{1}{\e}||\hat x_i||^2+c_1\sum_{i=1}^N ||\dot{z}_i||^2-c_3\nu V_0\\
	\leq &-\frac{1}{\e}||\hat x||^2+c_1||\dot{z}||^2 -c_3\nu V_0\\
\leq 	&-\frac{1}{\e}||\hat x||^2-(c_3\nu-c_1c_2)V_0
\end{align*}
Letting $c_3>\frac{c_1c_2+1}{\nu}$ gives 
\begin{align*}%\label{eq:thm1-eq1}
	\dot{\hat V} \leq &  -\frac{1}{\e}||\hat x||^2- V_0\triangleq \hat W(\hat x,\, \,\bar {\bf z})
\end{align*}

Recalling Theorem 2.1 in  \cite{krstic1995nonlinear}, we have that all trajectories of the closed-loop system \eqref{sys:composite-compact} are bounded over the time interval $[0,\,+\infty)$ and satisfy that $\lim_{t\to\infty} ||\hat x_i(t)||=0$. As immediate results, one can conclude the boundedness of $x_{j,i}(t)$ and $\theta(t)$.  Moreover, we can obtain that $\lim_{t\to\infty} \hat x_{1,i}(t)=0$, that is, $\lim_{t\to\infty}[y_i(t)-z_i(t)]=0$. Using the triangle inequality, we have  $|y_i(t)-y_i^*|\leq|y_i(t)-z_i(t)|+|z_i(t)-y_i^*|\to 0$ as $t\to +\infty$.  The proof is thus complete. % the LaSalle-Yoshizawa theorem  
\pe
	
\begin{remj}
		%\em
	Note that the considered agent \eqref{sys:agent} is high-order and subject to unknown dynamics, which includes both single and multiple integrators as its special cases. Thus, the theorem can be taken as an adaptive extension to existing results when the agent dynamics are exactly known \cite{gadjov2019passivity,romano2019dynamic}. Moreover, many typical actuating disturbances can be represented by the form \eqref{sys:agent} including the case when the disturbance is generated by a known autonomous linear dynamics as in \cite{romano2019dynamic,tang2020distributed}. Thus, we provide an alternative way to reject external disturbances other than the observer-based approach used in \cite{romano2019dynamic} and the internal model-based design in \cite{zhang2019distributed}.  
\end{remj}

%Particularly, when the analytical form of $f_i(\cdot)$ or $\nabla f_i(\cdot)$ is known to us, the optimal signal generator can be implemented independently. Following a similar proof, we can choose the gain parameter $\e$ as any positive constant to solve our problem. 
%	
%\begin{thm}\label{thm:main:offline}
%	Suppose Assumptions \ref{ass:system}--\ref{ass:graph} hold. Then, the distributed optimization problem determined by \eqref{sys:agent} and \eqref{opt:main} can be solved by the following control
%		\begin{align}\label{ctr:offline}
%		\begin{split}
%		u_i&=-\hat \theta_i^\top {\bm p}_{i}(x_{i},\,t)+\frac{1}{\e^{n_i}}[k_{1i}(x_{1,i}-z_i)+\sum\nolimits_{j=2}^{n_i}\e^{j-1}k_{ji}x_{j,\,i}]\\
%		\dot{\hat \theta}_i&={\bm p}_{i}(x_{i},\,t) b_{2i}^\top P_i \hat x_i\\
%		\dot{z}_i&=-\nabla f_i(z_i)-\sum\nolimits_{j=1}^N a_{ij}(\lm_i-\lm_j)\\
%		\dot{\lm}_i&=\sum\nolimits_{j=1}^N a_{ij}(z_i-r_j) 
%		\end{split}
%		\end{align} 
%	where the constant $k_{ji}$ is chosen as above and $\e>0$ is arbitrary for $i=1,\,\dots,\,N,\,j=1,\,\dots,\,n_i$.
%\end{thm}
	
\begin{remj}	%\em
	In the developed controller \eqref{ctr:offline}, we may choose $\phi_i(x_i,\,\hat \theta_i,\,z_i,\,t)=\Lambda_i{\bm p}_{i}(x_{i},\,t) b_{2i}^\top P_i \hat x_i$ with $\Lambda_{i}$ a chosen positive definite matrix. This matrix $\Lambda_i$ is called the adaption gain in the literature \cite{ioannou1995robust}. It can be used to achieve a fast adaption and then improve the transient performance of the controller to solve our Nash equilibrium seeking problem.
\end{remj}
	
%\begin{remj}
%	Without further information of the unknown dynamics, this controller may fail in practical applications if there are external disturbances or noises in measurements of $x_i$, although it is theoretically proved to achieve the optimization goal as $t$ goes to infinity. To tackle this problem, we can employ a $\sigma$-modification \cite{ioannou1995robust} for $\hat \theta_i$  with sacrificing some accuracy in control performance as follows:
%	\begin{align}\label{ctr:sigma}
%		\begin{split}
%		u_i&=-\hat \theta_i^\top {\bm p}_{i}(x_{i},\,t)+\frac{1}{\e^{n_i}}[k_{1i}(x_{1,i}-z_i)+\sum\nolimits_{j=2}^{n_i}\e^{j-1}k_{ji}x_{j,\,i}]\\
%		\dot{\hat \theta}_i&=-\sigma_{\theta_i}\hat\theta_i+{\bm p}_{i}(x_{i},\,t) b_{2i}^\top P_i \hat x_i\\
%		\dot{z}_i&=-\nabla f_i(y_i)-\sum\nolimits_{j=1}^N a_{ij}(\lm_i-\lm_j)\\
%		\dot{\lm}_i&=\sum\nolimits_{j=1}^N a_{ij}(z_i-r_j) 
%		\end{split}
%	\end{align} 
%		where $\sigma_{\theta_i}>0$ is a tunable parameter such that $\lim_{t\to+\infty}||y_i(t)-y^*||$ can be smaller than any desired positive constant.
%\end{remj}
%	
	
\subsection{Real-time Gradient Extension}

In the preceding  section, we implicitly assume the partial gradient function $\nabla_i J_i$ can be evaluated at any given estimate ${\bf z}^i$. This is the case when the analytic form of the local cost function is known by agent $i$.  However, in many circumstances, we may not have this knowledge and such partial gradient information can only be accessed or approximated when the real-time output strategy $y_i$ is taken. In this case, the generator \eqref{sys:generator} fails to be implemented. 

Let us replace $\nabla_i J_i({\bf z}^i)$ by $\nabla_i J_i(y_i,\,{\bf z}_{-i}^i)$ and obtain
\begin{align}\label{sys:generator-online}
\begin{split}
\dot{z}_i&=-\alpha \sum\nolimits_{j=1}^N a_{ij}({z}_i-{{z}}^j_i)-\nabla_i J_i(y_i,\,{\bf z}_{-i}^i)\\
\dot{z}_k^i&=-\alpha \sum\nolimits_{k=1}^N a_{ij}({{z}}_k^i-{{z}}_k^j),\quad k \in \mathcal{N}\backslash \{i\}
\end{split}
\end{align}
Although this dynamics is similar with \eqref{sys:generator}, it can not generate the expected Nash equilibrium by itself. In fact, denoting  $\Delta^1_i \triangleq  \nabla_i J_i({\bf z}^i)-\nabla_i J_i(y_i,\,{\bf z}_{-i}^i)$ gives
\begin{align*}
\begin{split}
\dot{z}_i&=-\alpha \sum\nolimits_{j=1}^N a_{ij}({z}_i-{{z}}^j_i)-\nabla_i J_i({\bf z}_{i})-\Delta_i^1\\
\dot{z}_k^i&=-\alpha \sum\nolimits_{k=1}^N a_{ij}({{z}}_k^i-{{z}}_k^j),\quad k \in \mathcal{N}\backslash \{i\}
\end{split}
\end{align*}
Compared with the optimal signal generator \eqref{sys:generator}, the error term $\Delta^1_i$ always exists except the case when $x_{1,\,i}=z_i$. Thus, there must be a discrepancy between ${\bm z}_i$ and $z^*$ when $\Delta^1_i\neq 0$.

Putting \eqref{sys:generator-compact-online}  into a compact form, we have
\begin{align}\label{sys:generator-compact-online}
\dot{{\bf z}}=-\alpha {\bf L}{\bf z}-R {\bm F}({\bf z})+R\Delta^1
\end{align}
where $\Delta^1=\mbox{col}(\Delta^1_1,\,\dots,\,\Delta^1_N)$. Note that system \eqref{sys:generator-compact-online} is exponentially stable when $\Delta^1\equiv{\bm 0}$ by Lemma \ref{lem:generator}. At the same time, one can verify that $\Delta_i^1$ is $l$-Lipschitz with respect to the tracking error $x_{1,\,i}-z_i$ (or $\tilde x_i$) for each $i\in \mathcal{N}$ by Assumption \ref{ass:lip-extended}. Recalling Lemma 4.6 in \cite{khalil2002nonlinear}, this system \eqref{sys:generator-compact-online} is roughly input-to-state stable with the tracking error $x_{1,\,i}-z_i $ as its input. This fact inspires us to develop fast tracking controllers for each agent to compensate the estimate error ${\bf z}_i-z^*$ and complete the whole design by decreasing the parameter $\e$.

To this end, we use the  same tracking controller for each agent as in the previous subsection. Jointly with the modified generator \eqref{sys:generator-online}, the overall controller to solve our problem  using only real-time gradients is presented as follows.
\begin{align}\label{ctr:online}
u_{i}&=-\hat \theta_i^\top {\bm p}_{i}(x_{i},\,t)+\frac{1}{\e^{n_i}}[k_{1i}(x_{1,i}-z_i)+\sum_{j=2}^{n_i}\e^{j-1}k_{ji}x_{j,\,i}] \nonumber\\
\dot{\hat \theta}_i&=\phi_i(x_i,\,\hat \theta_i,\, z_i,\,t) \nonumber\\
\dot{z}_i&=-\alpha \sum_{j=1}^N a_{ij}({z}_i-{{z}}^j_i)-\nabla_i J_i(y_i,\,{\bf z}_{-i}^i)\\
\dot{z}_k^i&=-\alpha \sum_{k=1}^N a_{ij}({{z}}_k^i-{{z}}_k^j),\quad k \in \mathcal{N}\backslash  \{i\} \nonumber
\end{align}
with $\e>0$ to be specified later.

Here is a theorem to ensure the effectiveness of this controller to solve our problem.	
\begin{thmj}\label{thm:main-online}
	%\em
	Suppose Assumptions \ref{ass:convex+monotone}--\ref{ass:lip-extended} hold. Choose  $\alpha$, $k_{1i}$, $\dots$  $k_{n_ii}$, and $\phi_i$ as in Theorem \ref{thm:main} for each $i\in \mathcal{N}$. Then, there exists a constant $\e^*>0$ such that the Nash equilibrium seeking problem for multi-agent system \eqref{sys:agent} is solved by distributed controllers of the form \eqref{ctr:online} for any $\e\in (0,\,\e^*)$.
\end{thmj}
\pb  Under the new controller \eqref{ctr:online}, the whole composite system is then 
\begin{align*}
\dot{x}_{1,\,i}&=x_{2, i} \\ 
\vdots&\\
\dot{x}_{n_i,i}&=(\theta_i^\top-\hat \theta_i^\top) {\bm p}_{i}(x_{i},\,t)+\frac{1}{\e^{n_i}}[k_{1i}(x_{1,i}-z_i)\\
& +\sum\nolimits_{j=2}^{n_i}\e^{j-1}k_{ji}x_{j,\,i}]\\ 
\dot{\hat \theta}_i&=\phi_i(x_i,\,\hat \theta_i,\,z_i,\,t)\\
\dot{z}_i&=-\alpha \sum_{j=1}^N a_{ij}({z}_i-{{z}}^j_i)-\nabla_i J_i(y_i,\,{\bf z}_{-i}^i)\\
\dot{z}_k^i&=-\alpha \sum_{k=1}^N a_{ij}({{z}}_k^i-{{z}}_k^j),\quad k \in \mathcal{N}\backslash  \{i\} \nonumber
\end{align*}

By applying the same transformation of coordinates as in system \eqref{sys:closed-spform}, we can put the above composite system into a compact form as follows.
\begin{align}\label{sys:composite-compact-online}
\begin{split}
\dot{\hat x}&=\frac{1}{\e}A\hat  x-B_{1} \dot{z}+EB_{2} {\bm p}^\top (x,\,t) \bar \theta\\
\dot{\bar {\theta}}&=\phi(x,\,\hat \theta,\,z,\,t) \\
\dot{{\bf z}}&=-\alpha {\bf L}{\bf z}-R {\bm F}({\bf z})- R\Delta^1
\end{split}
\end{align}
Note that the third subsystem can be further rewritten in the $(\bar {\bf z}_1,\,{\bf \bar z}_2)$ coordinate as 
\begin{align*}
\dot{\bar{\bf z}}_1&=-(M_1^\top\otimes I_N)R \Delta -(M_1^\top\otimes I_N)R \Delta^1\\
\dot{\bar {\bf z}}_2&=-\alpha [(M_2^\top L M_2)\otimes I_{N}]\bar {\bf z}_2-(M_2^\top\otimes I_N) R\Delta-(M_2^\top\otimes I_N) R\Delta^1
\end{align*}

Next, we use the same Lyapunov function $\hat V=\sum_{i=1}^N \hat V_i+c_3V_0$  for this new composite system \eqref{sys:composite-compact} with $c_4>0$ defined as in the proof of Theorem \ref{thm:main}. By similar arguments as in the proof of Theorem \ref{thm:main}, we have
\begin{align*}%\label{eq:thm1-eq1}
\dot{\hat V} &\leq -\sum_{i=1}^N \frac{1}{\e}||\hat x_i||^2+c_1\sum_{i=1}^N ||\dot{z}_i||^2-c_3 {\bar{\bf z}}_1^\top (M_1^\top\otimes I_N)R (\Delta+\Delta^1)\\
&- c_3 {\bar{\bf z}}_2^\top\{\alpha [(M_2^\top L M_2)\otimes I_{N}]\bar {\bf z}_2+(M_2^\top\otimes I_N) R(\Delta+\Delta^1)\}\\
&\leq -\sum_{i=1}^N \frac{1}{\e}||\hat x_i||^2+c_1\sum_{i=1}^N ||\dot{z}_i||^2-c_3\nu V_0 -c_3{\tilde {\bf z}}^\top R \Delta^1\\
&\leq -\frac{1}{\e}||\hat x||^2- V_0- c_3||{\bar{\bf z}}|||| \Delta^1||
\end{align*}
where we use $||{\tilde {\bf z}}||=||{\bar{\bf z}}||$. By Young's inequality and the $l$-Lipschitzness of $\Delta^1$ with respect to $\hat x$, it follows that
\begin{align*}%\label{eq:thm1-eq1}
\dot{\hat V}&  \leq  -\frac{1}{\e}||\hat x||^2- V_0+\frac{1}{4}||{\bar{\bf z}}||^2+ 4c_3^2l^2||\hat x||^2\\
& \leq  -(\frac{1}{\e}-4c_3^2l^2 )||\hat x||^2- \frac{1}{2}V_0 
\end{align*}
Setting $\e^*=\frac{1}{4c_3^2l^2+1}$ and $\e \in (0,\,\e^*)$, we can obtain  the following inequality:
\begin{align*}%\label{eq:thm1-eq1}
\dot{\hat V}& \leq  - ||\hat x||^2- \frac{1}{2}V_0 
\end{align*}
At this moment, we recall Theorem 2.1 in  \cite{krstic1995nonlinear} again and conclude that all trajectories of the closed-loop system \eqref{sys:composite-compact-online} are bounded over the time interval $[0,\,+\infty)$ and satisfy that $\lim_{t\to\infty} ||\hat x_i(t)||=0$ and $\lim_{t\to\infty} V_0(t)=0$. Then, we confirm the boundedness of $x_{j,i}(t)$, $\theta(t)$ and conclude $\lim_{t\to\infty}[y_i(t)-z_i(t)]=0$, $\lim_{t\to\infty}[z_i(t)-y^*]=0$. By the triangle inequality again, it follows that $|y_i(t)-y_i^*|\leq|y_i(t)-z_i(t)|+|z_i(t)-y_i^*|\to 0$ as $t\to +\infty$.  The proof is thus complete. % the LaSalle-Yoshizawa theorem  
\pe

\begin{remj}
	Note that the given choice of $\e^*$ heavily relies on some bounds of matrix norms. Thus, determining the largest parameter $\e^*$ in controller \eqref{ctr:online} might be nontrivial. In practice, one may choose an applicable parameter  $\e$ by numerical simulations to avoid this tedious job. 
\end{remj}

\subsection{Parameter Convergence}

From the proofs of Theorems \ref{thm:main} and \ref{thm:main-online}, one can merely conclude that  $\hat \theta_i(t)$ converges to some constant as $t$ tends to $+\infty$. However, this constant may not be the associated true value $\theta_i$. Since parameter convergence has been shown to be essential in achieving robustness of the adaptive controllers \cite{ioannou1995robust,mazenc2009uniform}, we assert conditions under which the estimator $\hat \theta_i(t)$ will converge to its true value $\theta_i$ as $t$ tends to $+\infty$. 
	
To this end, we further assume the basis function ${\bm p}_i(x_i,\,t)$ satisfying the following condition.
	\begin{assj}\label{ass:PE}
		For any $i=1,\,\dots,\,N$, along the trajectory of the closed-loop system composed of \eqref{sys:agent} and \eqref{ctr:offline}, there exist positive constants $m$,\,$t_0$,\,$T_0$ such that the function ${{\bm p}}_i(x_i(t),\,t)$ is uniformly bounded and the following inequality is satisfied:
		\begin{align*}%\label{eq:PE}
		\frac{1}{T_0}\int_{t}^{t+T_0}{\bm p}_i(x_i(\tau),\,\tau){\bm p}^\top(x_i(\tau),\,\tau){\rm d}\tau \geq mI_{n_{\theta_i}},\, \forall t\geq t_0
		\end{align*}
	\end{assj}
	
Note that $x_i(t)$ is ultimately bounded by Theorem \ref{thm:main}, the boundedness of ${\bm p}_i(x_i(t),\,t)$ is not too strict. The above inequality is known as a version of the well-known persistence of excitation (PE) condition and  has widely used in adaptive control literature \cite{krstic1995nonlinear,hu2014adaptive,chen2015stabilization}. 
	
\begin{thmj}\label{thm:parameter}	%\em
	Suppose Assumptions \ref{ass:convex+monotone}--\ref{ass:PE} hold. Then, along the trajectory of system \eqref{sys:agent} under the controller \eqref{ctr:offline}, it holds that $\lim_{t\to+\infty}\hat \theta_i(t)=\theta_i$ for  $i=1,\,\dots,\,N$.
\end{thmj}
\pb To show this theorem, we first claim that $\lim_{t\to+\infty}\bar \theta_i^\top(t){\bm p}_i(x_i(t),\,t)=0$.  By the proof of Theorem \ref{thm:main}, we have $\hat x_i(\infty)=\int_{0}^{+\infty}\dot{\hat x}_i(\tau){\rm d}\tau=0$. From the uniform boundedness of associated variables and Assumption \ref{ass:PE}, it follows that $\ddot{\hat x}_i(t)$ is also bounded. Using Lemma 8.2 in \cite{khalil2002nonlinear} to $\dot{\hat x}_i(t)$ implies that $\lim_{t\to+\infty}\dot{\hat x}_i(t)=0$, which confirms this claim.
	
Next, since $\dot{\bar \theta}_i=\dot{\hat \theta}_i={\bm p}_{i}(x_{i},\,t) b_{2i}^\top P_i \hat x_i$, it follows that $\lim_{t\to+\infty}\dot{\bar \theta}_i(t)=0$ since $\lim_{t\to+\infty}\dot{\hat x}_i(t)=0$. According to Lemma 1 in  \cite{ortega1993asymptotic}, the two facts $\lim_{t\to+\infty}\dot{\bar \theta}_i(t)=0$ and $\lim_{t\to+\infty}\bar \theta_i^\top(t){\bm p}_i(x_i(t),\,t)=0$ provide us that $\lim_{t\to+\infty}\bar \theta_i(t)=0$ under Assumption \ref{ass:PE}. The proof is thus complete. 
\pe

\begin{remj}\label{rem:parameter}	%\em
	Note that the unknown dynamics is supposed to be linearly parameterized  in this paper.  This structure allows us to further improve this theorem and apply it to any number of components in ${\bm p}_i(x_i,\,t)$ satisfying such a PE condition. In this way, we  can address the parameter convergence problem in a more precise way. Specially, when the basis function is time-invariant, the $j$-th component ${\bm p}_{j,i}(x_i)$ of ${\bm p}_i(x_i)$ is persistently excited if $\lim_{x_i\to \mbox{col}(y_i^*,\,0,\,\dots,\,0)}{\bm p}_{j,i}(x_i)\neq 0$, which further ensures the convergence of $\hat \theta_{j,i}(t)$ to $\theta_{j,i}$ as $t$ goes to infinity.	
\end{remj}

\section{Simulation}\label{sec:simu}  %%Real time gradient
	
In this section, we present numerical examples to illustrate the effectiveness of our preceding design.  

% Example 1: Optical + Disturbance
% Example 2: Robotics
% Example 3: van Der Pol
\begin{figure}
	\centering
	%	\subfigure[]{
	%	\centering
	\begin{tikzpicture}[scale=1, shorten >=1pt, node distance=1.5cm, >=stealth',
	every state/.style ={circle, minimum width=0.7cm, minimum height=0.7cm}]
	\node[align=center,state](node1) {\scriptsize 1};
	\node[align=center,state](node2)[right of=node1]{\scriptsize 2};
	\node[align=center,state](node3)[right of=node2]{\scriptsize 3};
	\node[align=center,state](node4)[right of=node3]{\scriptsize 4};
	\node[align=center,state](node5)[right of=node4]{\scriptsize 5};
	\node[align=center,state](node6)[below of=node5]{\scriptsize 6};
	\node[align=center,state](node7)[left of=node6]{\scriptsize 3};
	\node[align=center,state](node8)[left of=node7]{\scriptsize 8};
	\node[align=center,state](node9)[left of=node8]{\scriptsize 9};
	\node[align=center,state](node10)[left of=node9]{\scriptsize 10};
	\path[<-]  
	(node1) edge (node2)
	(node2) edge (node3)
	(node3) edge (node4)
	(node4) edge (node5)
	(node5) edge (node6)
	(node6) edge (node7)
	(node7) edge (node8)
	(node8) edge (node9)
	(node9) edge (node10)
	(node10) edge (node1);
	\end{tikzpicture}	
	%	}
	%	\subfigure[]{
	%		\includegraphics[width=0.48\textwidth]{opt-fig-journal-digraph-adaptive.eps}\label{fig:simu1}
	%	}
	\caption{Communication digraph $\mathcal G$ in Example I.}\label{fig:graph-exam-level}
\end{figure}

\begin{figure}
	\centering
	\includegraphics[width=0.4\textwidth]{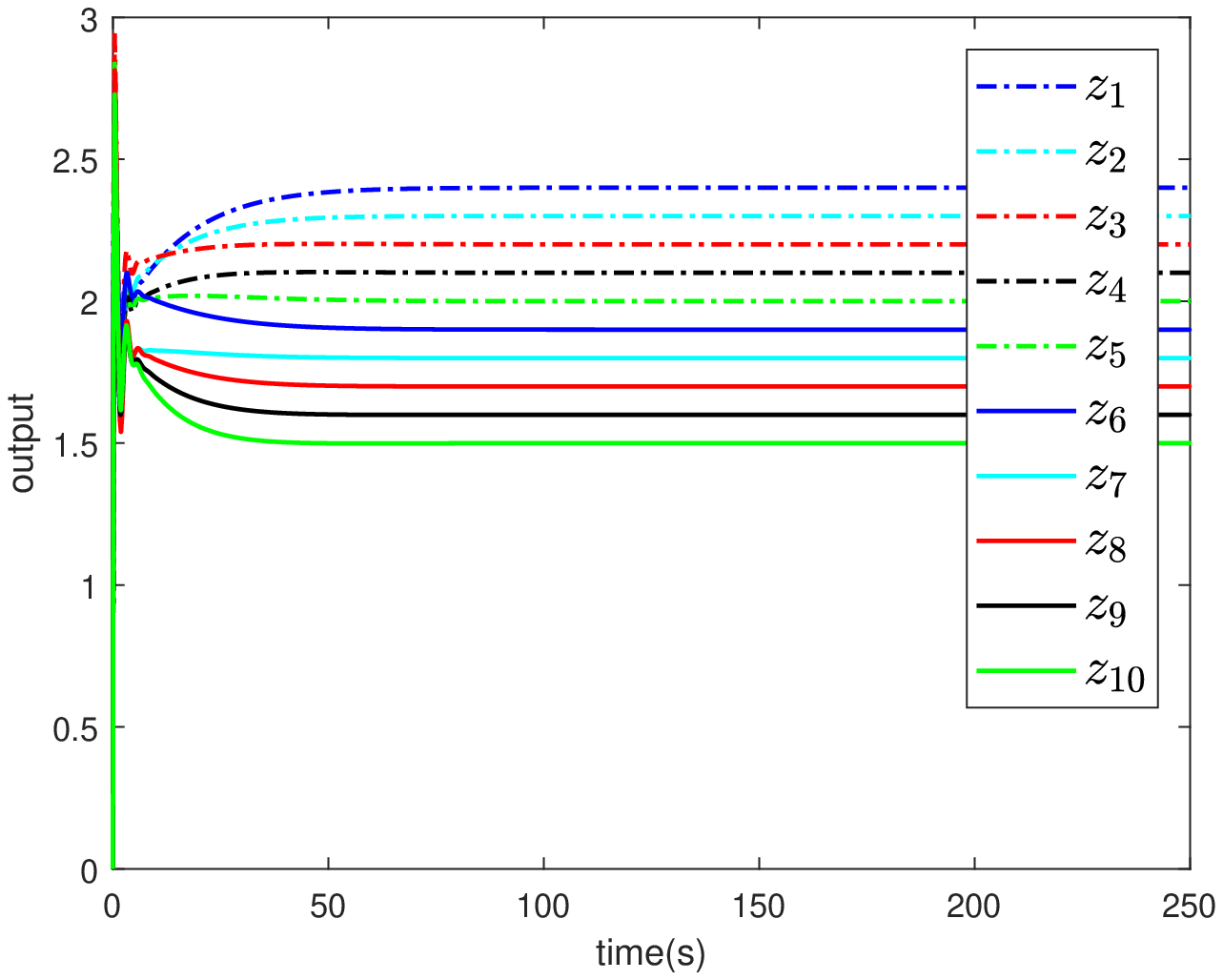}
	\caption{Profiles of $z_i(t)$ under the  generator \eqref{sys:generator}.}\label{fig:simu-exam-level-reference}
\end{figure}

{\it Example 1}: Consider a group of $N=10$ firms and suppose they produce a homogeneous perishable commodity \cite{raafat1991survey}. The inventory system at firm $i$  can be modeled as
\begin{align}\label{agent:exam-level}
	\dot{I}_i=-\gamma_iI_i+P_i-D_i,\quad  i\in \mathcal{N}
\end{align}
where $I_i$ is the inventory level, $\gamma_{i}$ is the deterioration rate, $P_i$ is the production rate, and $D_i$ is demand rate at firm $i$. Suppose these firms can share information through a cycle digraph with unity weights depicted as Fig.~\ref{fig:graph-exam-level}.

To meet a safety requirement imposed by some authority in this market (e.g., the government), these firms are expected to maintain their total inventory at certain level $I_r>0$. The total cost function of firm $i$ is given as  $J_i(I_i,\,I_{-i})=C_i(I_i)-I_i*\sigma(I_1,\,\dots,\,I_N)$, where $C_i(s)=\alpha_is$ is the storage cost and $\sigma(I_1,\,\dots,\,I_N)=\delta_0(I_r-\sum_{i=1}^N I_i)$ is the subsidyis per unit provided by this market authority with known constants $\alpha_i,\delta_0>0$. To make it more interesting, we suppose that the deterioration rate $\gamma_i$ and the demand rate $D_i$ at agent $i$ are both constant but unknown. Letting  $\theta_i=\mbox{col}(-\gamma_i,\,D_i)$ and ${\bm p}(I_i,\,t)=\mbox{col}(-I_i,\,1)$, we can find that these firms play a noncooperative game with cost function $J_i$ and unknown dynamics of the form  \eqref{agent:exam-level}. Moreover, Assumptions \ref{ass:convex+monotone}--\ref{ass:lip-extended} can be practically verified. Then, according to Theorem \ref{thm:main}, we can determine a distributed controller of the form \eqref{ctr:offline} with $n_i=1$ to solve the formulated problem for agent \eqref{agent:exam-level}.

\begin{figure}
	\centering
	\includegraphics[width=0.4\textwidth]{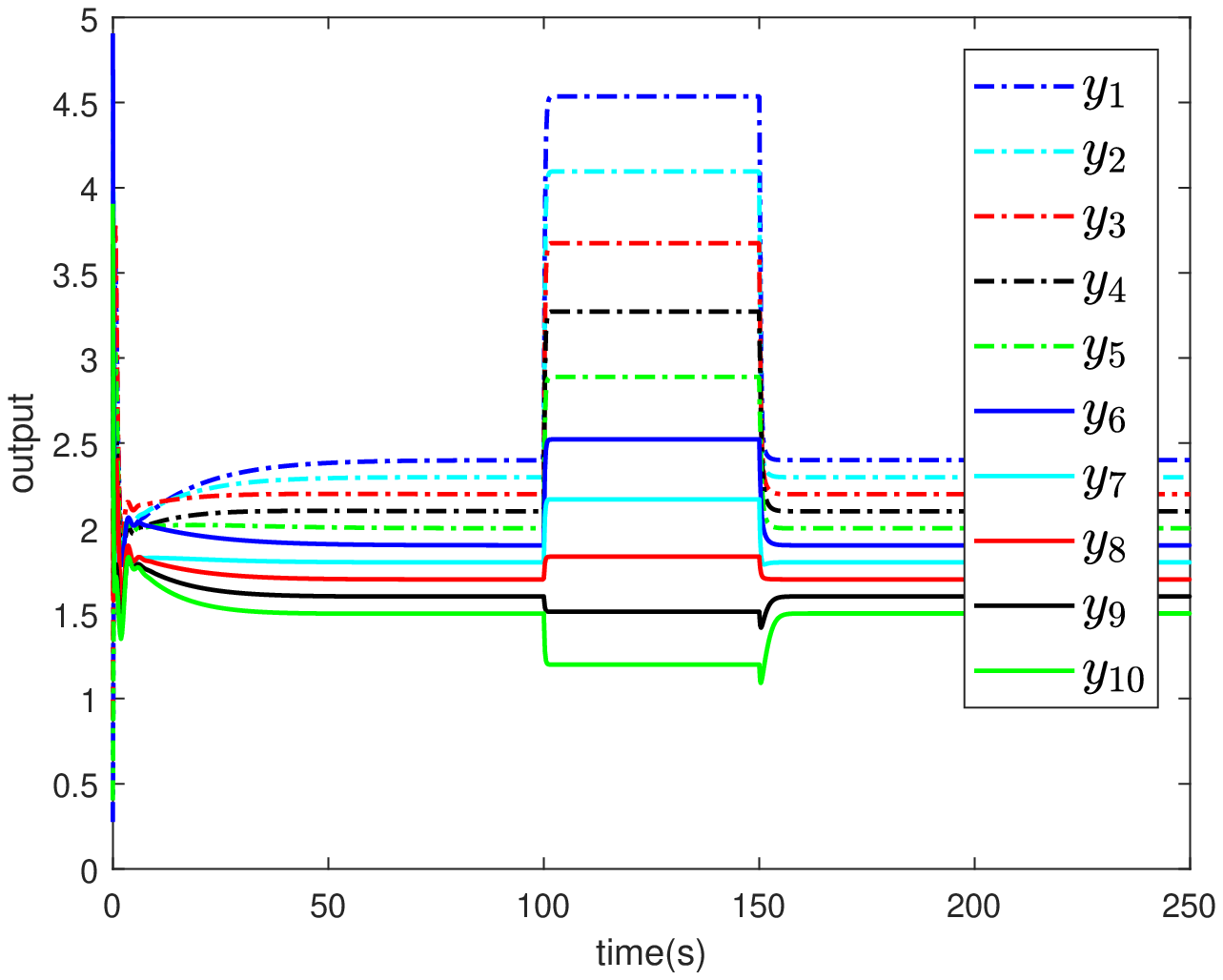}
	\caption{Profiles of $y_i(t)$ under the controller \eqref{ctr:offline}.}\label{fig:simu-exam-level-output}
\end{figure}

\begin{figure}
	\centering
	\includegraphics[width=0.4\textwidth]{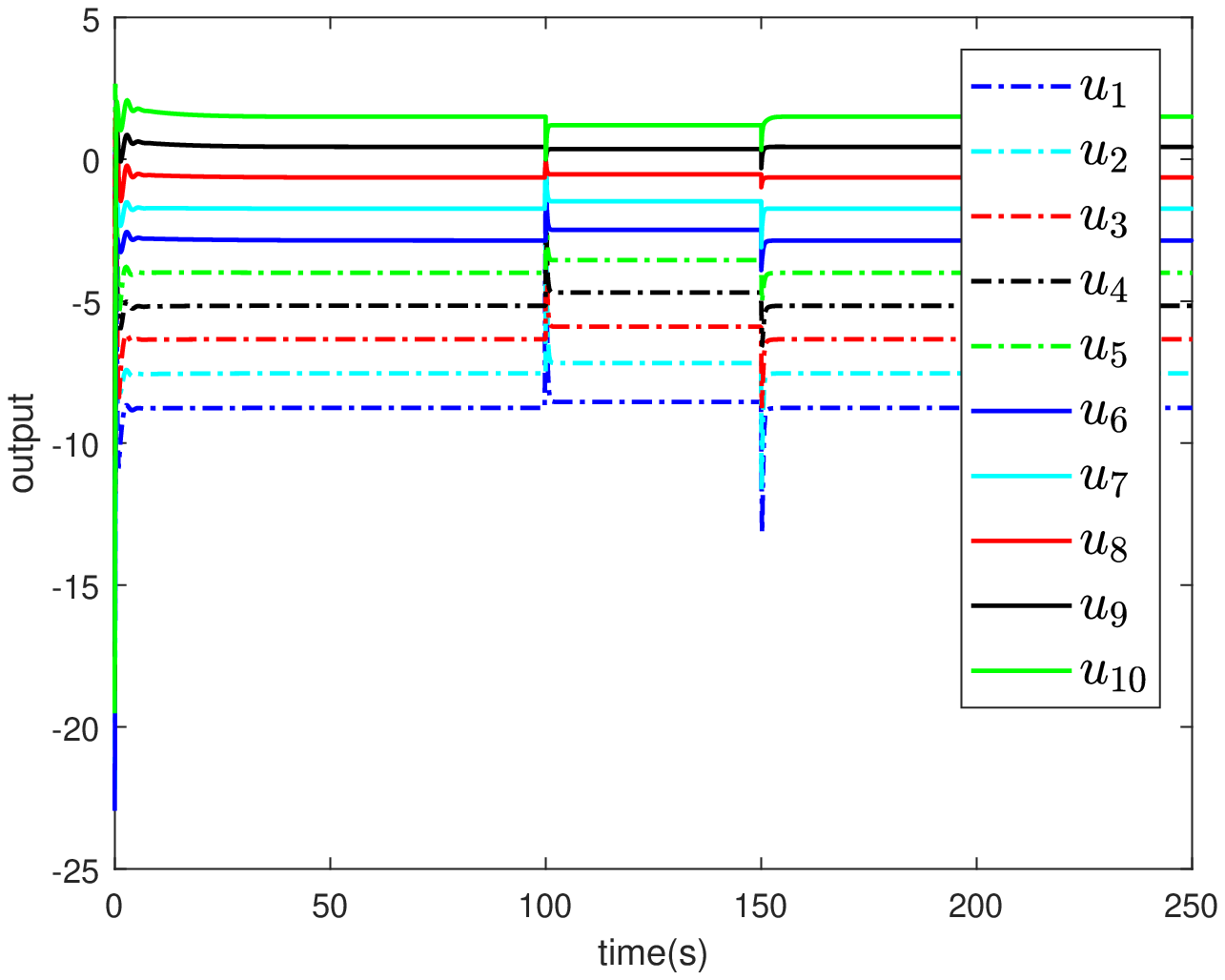}
	\caption{Profiles of $u_i(t)$ under the controller \eqref{ctr:offline}.}\label{fig:simu-exam-level-control}
\end{figure}

For simulations, we assume $N=10$ and let $\alpha_i=i/10$, $I_r=22$, $\delta_0=1$, $\theta_i=\frac{i}{2}$, and $D_i={10-i}$. The Nash equilibrium  can be determined as $y^*=\mbox{col}(y_1^*,\,\dots,\,y_{10})$ with $y_i^*=2.5-\alpha_i$ for $i=1,\,\dots,\,N$.  Choose $\alpha=4$, $k_1=-4$ for the controller \eqref{ctr:offline} and the initial inventory levels randomly from $[0,\,5]$. The generated reference for each agent is depicted as Fig.\ref{fig:simu-exam-level-reference}.  To verify the effectiveness of our controller to compensate the unknown dynamics, we shut down the adaptive part between $t=100{\rm s}$ and $t=150 {\rm s}$. The profiles of agent outputs and control efforts are shown in Figs.~\ref{fig:simu-exam-level-output} and \ref{fig:simu-exam-level-control}, where the expected Nash equilibrium $y^*$ is quickly reached before $t=100 {\rm s}$ and the control efforts are maintained to be bounded. Moreover, we can find that the steady-state of each agent deviates from the expected Nash equilibrium $y^*$ after $t=100 {\rm s}$ and soon recovers after $t=150 {\rm s}$.

\begin{figure}[!htp]
	\centering
	%	\subfigure[]{
	%	\centering
	\begin{tikzpicture}[scale=1, shorten >=1pt, node distance=1.5cm, >=stealth',
	every state/.style ={circle, minimum width=0.7cm, minimum height=0.7cm}]
	\node[align=center,state](node1) {\scriptsize 1};
	\node[align=center,state](node5)[right of=node1]{\scriptsize 5};
	\node[align=center](node53)[right of=node5]{};
	\node[align=center,state](node3)[right of=node53]{\scriptsize 3};
	\node[align=center,state](node2)[right of=node3]{\scriptsize 2};
	\node[align=center,state](node4)[above of=node53]{\scriptsize 4};
	\path[<->]  
	(node1) edge (node5)
	(node5) edge (node3)
	(node3) edge (node2)
	(node5) edge (node4)
	(node3) edge (node4)
	;
	\end{tikzpicture}	
	%	}
	%	\subfigure[]{
	%		\includegraphics[width=0.48\textwidth]{opt-fig-journal-digraph-adaptive.eps}\label{fig:simu1}
	%	}
	\caption{Communication digraph $\mathcal G$ in Example 2.}\label{fig:graph-exam-sensor}
\end{figure}
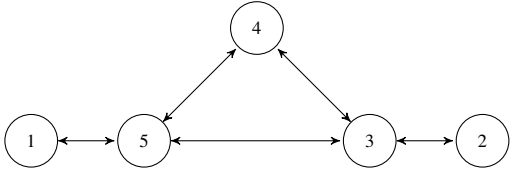

{\it  Example 2}: Consider the sensor network example discussed in \cite{romano2019dynamic}. Suppose we have a group of five force-actuated mobile robots in the plane modeled as follows:
\begin{align}\label{agent:exam-sensor}
\ddot{y}_i=u_i+d_i
\end{align}
where $y_i\in \R^2$, $\dot{y}_i\in \R^2$, and $u_i\in \R^2$ are the position, velocity, and control input of agent $i$.  Here $d_i\in \R^2$ is a local actuating disturbance of agent $i$. Similar as in \cite{romano2019dynamic}, agent $i$ is supposed to have a local cost function depending upon the positions of all robots $(y_i,\,y_{-i})$ as follows:
\begin{align*}
J_i(y_i,\,y_{-i})=y_i^\top y_i+y_i^\top r_i+ \sum_{j=1}^5||y_i-y_j||^2
\end{align*}
where $r_1=\mbox{col}(2,\,-2)$, $r_2=\mbox{col}(-2,\,-2)$, $r_3=\mbox{col}(-4,\,2)$, $r_4=\mbox{col}(2,\,-4)$, and $r_5=\mbox{col}(3,\,3)$.  The communication topology is represented by an undirected graph with unity edge weights depicted as Fig.~\ref{fig:graph-exam-sensor}. Then,  these agents play a noncooperative game under a partial information scenario. Moreover, Assumptions \ref{ass:convex+monotone}--\ref{ass:lip-extended} can be verified and the Nash equilibrium is determined as $y^*=\mbox{col}(y_1^*,\,\dots,\,y_5^*)$ with $y_i^*=-\frac{r_i+\sum_{j=1}^5 r_j}{12}$.  

\begin{figure}[!htp]
	\centering
	\includegraphics[width=0.4\textwidth]{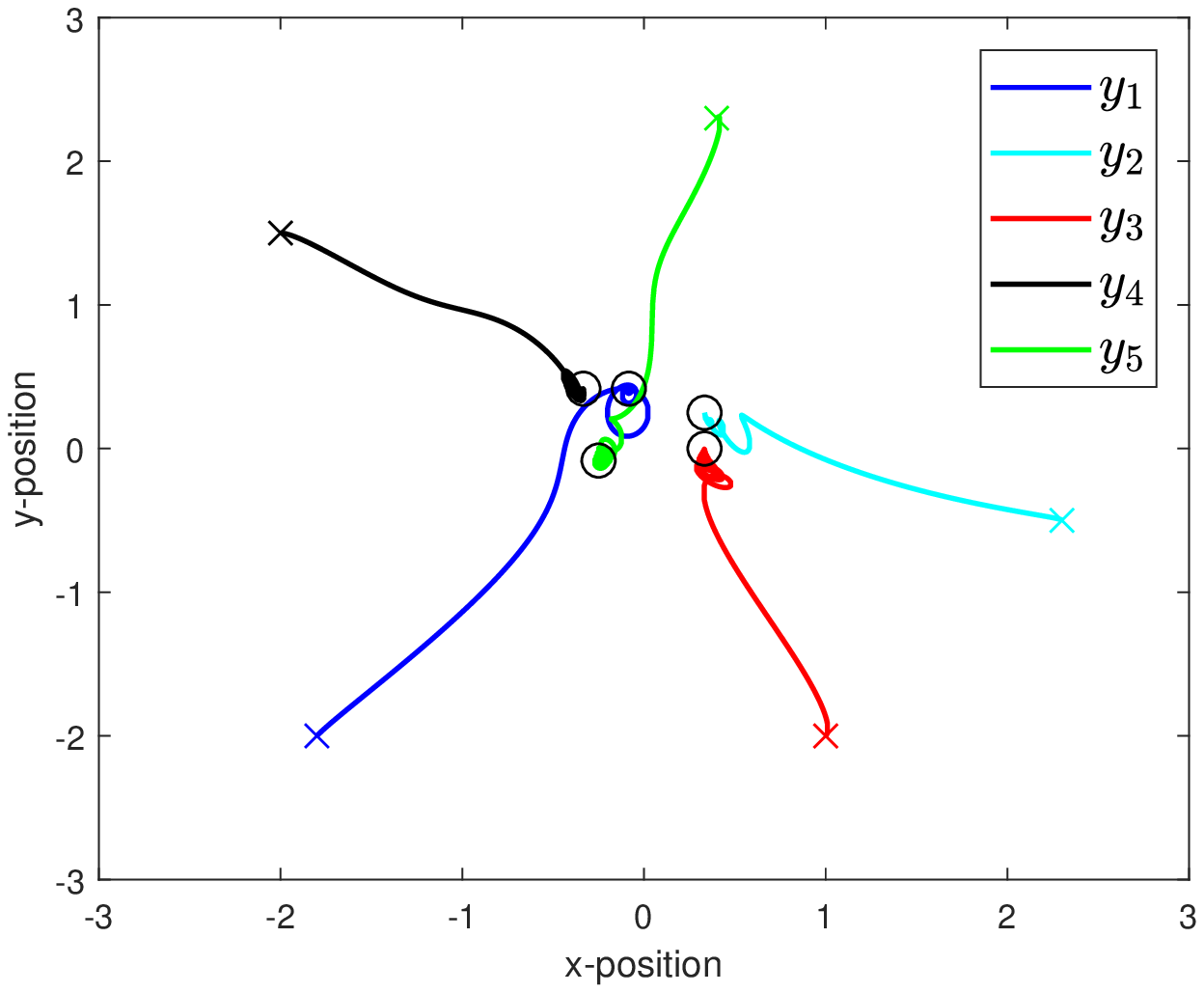}
	\caption{Profiles of $y_i(t)$ under the controller \eqref{ctr:exam-sensor}.}\label{fig:simu-exam-sensor-plane}
\end{figure}

\begin{figure}[!htp]
	\centering
	\includegraphics[width=0.4\textwidth]{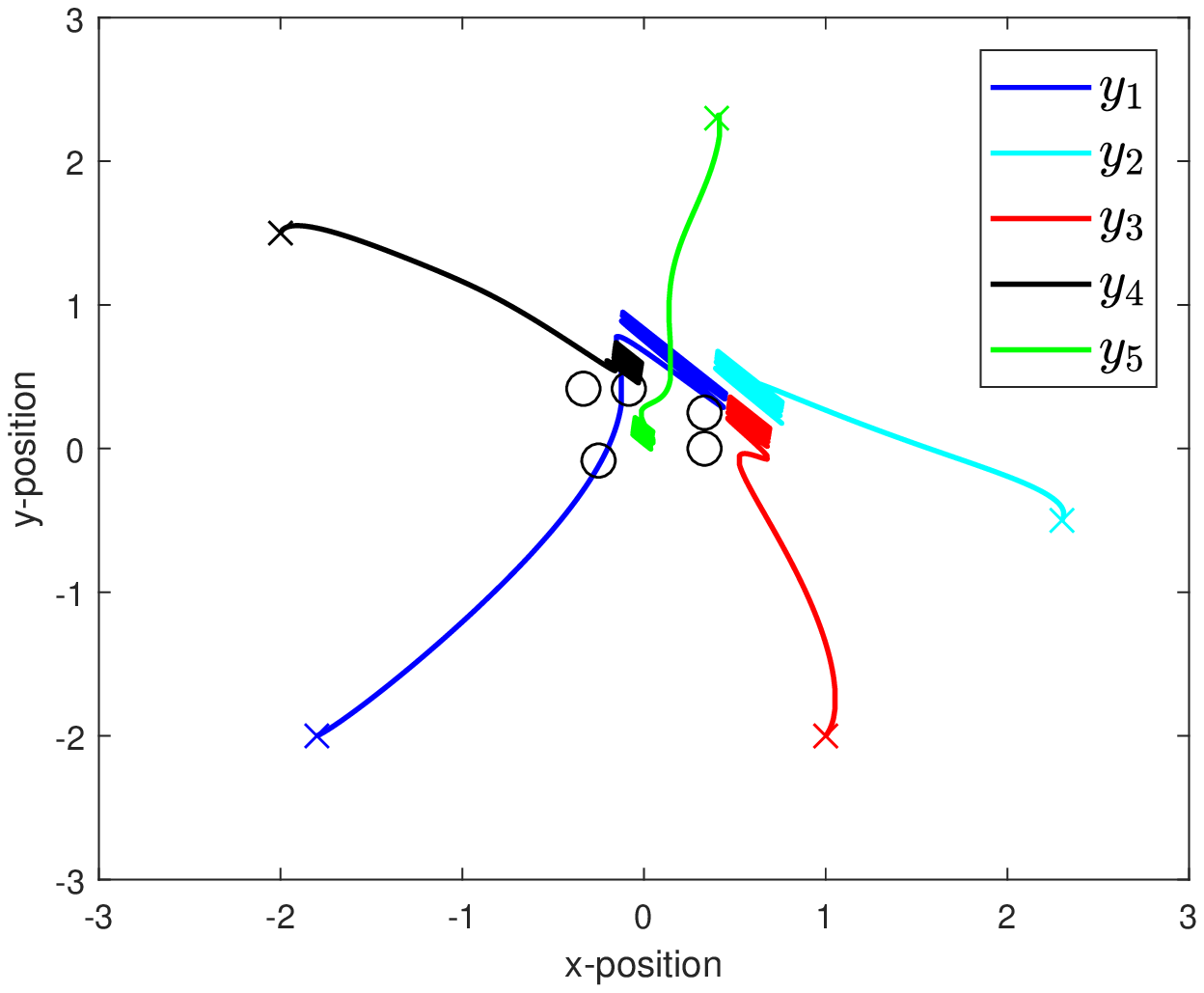}
	\caption{Profiles of $y_i(t)$ under the controller \eqref{ctr:exam-sensor-d}.}\label{fig:simu-exam-sensor-plane-d}
\end{figure}

Different from the simulation in \cite{romano2019dynamic}, agent $i$ is supposed to have a nonconstant actuating disturbance modeled by $d_i(t)=D_i v_i(t),\, \dot{v}_i=S_iv_i$ with 
$$D_i=\begin{bmatrix}
1+\mu_1&1+\mu_2&\mu_3\\
1+\mu_4&\mu_5&1+\mu_6\\
\end{bmatrix},\quad S_i=\begin{bmatrix}
0&0&0\\
0&0&i\\
0&-i&0
\end{bmatrix}$$
where ${\bm \mu}=\mbox{col}(\mu_1,\,\dots,\,\mu_6) \in \R^6$ is an uncertain parameter vector satisfying $|\mu_i|\leq 0.2$.  Although the pair $(D,\,S)$ is verified to be observable, the observer-based approach proposed in \cite{romano2019dynamic} fails to solve this problem due to the uncertain parameter ${\bm \mu}$.   Next, we show how to solve it by choosing a distributed controller of the form \eqref{ctr:offline} for each input channel of \eqref{agent:exam-sensor}.  

We resort to the fact that the disturbance can be represented as $d_i(t)=A_{0i}+A_{1i}\sin(i t)+A_{2i}\cos(i t)$ for some constant vectors $A_{0i},\, A_{1i},\, A_{2i}\in \R^2$ depending upon the initial value $v_i(0)$ and matrix $D$.  Then agent \eqref{agent:exam-sensor} can be rewritten into the form \eqref{sys:agent} with $n_i=2$, $x_{1,\,i}=y_i$, $x_{2,\,i}=\dot{y}_i$, ${\bm p}_i(x_i,\,t)=\mbox{col}(1,\,\sin(i t),\,\cos( i t))$, and $\theta_i=\mbox{col}(A_{0i}^\top,\, A_{1i}^\top,\, A_{2i}^\top)\in \R^{3\times 2}$. Then, we present the following controller for agent $i$:
\begin{align}\label{ctr:exam-sensor}
\begin{split}
u_i&=-\hat \theta^\top_{i}{\bm p}_i(x_i,\,t)-4(x_{1,\,i}-z_{1,\,i})-4x_{2,\,i}\\
\dot{\hat \theta}_i&=5 {\bm p}_i(x_i,\,t)[\frac{1}{4}(x_{1,\,i}  -z_i )+\frac{5}{16}x_{2,\,i}]^\top
\end{split}
\end{align}

For simulations, we choose the initial conditions randomly and mark the start position of each robot by crosses. The evolution of robots' positions is shown in Fig.~\ref{fig:simu-exam-sensor-plane}. One can find that the robots finally reach the Nash equilibrium position of the associated noncooperative game marked by circles. In comparison, we remove the adaption component in \eqref{ctr:exam-sensor} and use the following static controller for agent $i$: 
\begin{align}\label{ctr:exam-sensor-d}
\begin{split}
u_i&=-4(x_{1,\,i}-z_{1,\,i})-4x_{2,\,i}
\end{split}
\end{align}
Note that the closed-loop system under this controller is input-to-state stable with respect to the actuating disturbance as the input. Then, the output of agents will finally enter into a neighborhood of the Nash equilibrium, whose size depends on the strength of the disturbance. Moreover, these outputs can not converge to the expected position as shown in Fig.~\ref{fig:simu-exam-sensor-plane-d}. These observations verify the effectiveness of our controller in handing unknown external disturbances.

\begin{figure}
	\centering
	%	\subfigure[]{
	%	\centering
	\begin{tikzpicture}[scale=1, shorten >=1pt, node distance=1.8cm, >=stealth',
	every state/.style ={circle, minimum width=0.7cm, minimum height=0.7cm}]
		\node[align=center,state](node1) {\scriptsize 1};
		\node[align=center,state](node2)[right of=node1]{\scriptsize 2};
		\node[align=center,state](node3)[right of=node2]{\scriptsize 3};
		\node[align=center,state](node4)[right of=node3]{\scriptsize 4};
		\path[->]  (node1) edge (node2)
		(node2) edge (node3)
		(node3) edge (node4)
		(node4) edge [bend right=35]  (node1);
		\path[<->] (node1) edge [bend right=35] (node3)
		(node2) edge [bend right=35]  (node4)
		;
	\end{tikzpicture}	
	%	}
	%	\subfigure[]{
	%		\includegraphics[width=0.48\textwidth]{opt-fig-journal-digraph-adaptive.eps}\label{fig:simu1}
	%	}
	\caption{Communication digraph $\mathcal G$ in Example 2.}\label{fig:graph-exam-vandepol}
\end{figure}
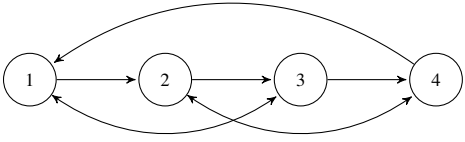

{\it  Example 3}:  Consider a multi-agent system including four controlled Van der Pol systems as follows.
	\begin{align*}
	\dot{x}_{1,i}&=x_{2, i}\\
	\dot{x}_{2,i}&= -a_i x_{1,i}+b_i (1-x_{1,i}^2)x_{2,i}+u_i\\
	y_i&=x_{1,i},\quad i=1,\,2,\,3,\,4
	\end{align*}
where $a_i,\,b_i$ are unknown positive constants. % The trajectories of the unforced system with different initial conditions when $a_i=b_i=1$ are depicted in Figure \ref{fig:phase}. 
The information sharing graph of this multi-agent system is depicted in Fig.~\ref{fig:graph-exam-vandepol} with unity edge weights.

We consider the Nash equilibrium seeking problem for this multi-agent system with a local cost function $J_i(y_i,\,y_{-i})=(y_i-y_{i0})^2-y_i(p_i\sum_{i=1}^4 y_i+q_i)$ for agent $i$ with $i=1,\, 2,\,3,\,4$. Note that all these agents have unknown nonlinear dynamics. To make it more interesting, we also assume that agent $i$ has an actuating disturbance $d_i(t)$ as in {\rm  Example 2} but with different system matrices
\begin{align*}
D_i=\begin{bmatrix}
1+\mu_1~~\mu_2
\end{bmatrix},\quad S_i=\begin{bmatrix}
0&i\\
-i&0
\end{bmatrix}
\end{align*}
with uncertain parameters $|\mu_1|\leq 0.5$ and $|\mu_2|\leq 0.5$.
	
Denote $\Delta_i(x_i,\,t)=-a_i x_{1,i}+b_i (1-x_{1,i}^2)x_{2,i}+d_i(t)$.  Note that $d_i(t)=A_{1i}\sin(i t)+A_{2i}\cos(i t)$ for some constants $A_{1i}$,  $A_{2i}$ depending upon the initial value $v_i(0)$ and $D_i$. We let 
\begin{align*}
&{\bm p}_i(x_i,\,t)=\mbox{col}(-x_{1,i},\,(1-x_{1,i}^2)x_{2,i},\,\sin(i t),\,\cos( i t))\\
&\theta_i=\mbox{col}(\theta_{1,i},\,\dots,\,\theta_{4,i})=\mbox{col}(a_i,\,b_i,\,A_{1i},\,A_{2i})
\end{align*}
Then, these agents have been put into the form \eqref{sys:agent} with basis function ${\bm p}_i(x_i,\,t)$ defined as above and an unknown parameter vector $\theta_i\in \R^4$. 

\begin{figure}
	\centering
	\includegraphics[width=0.4\textwidth]{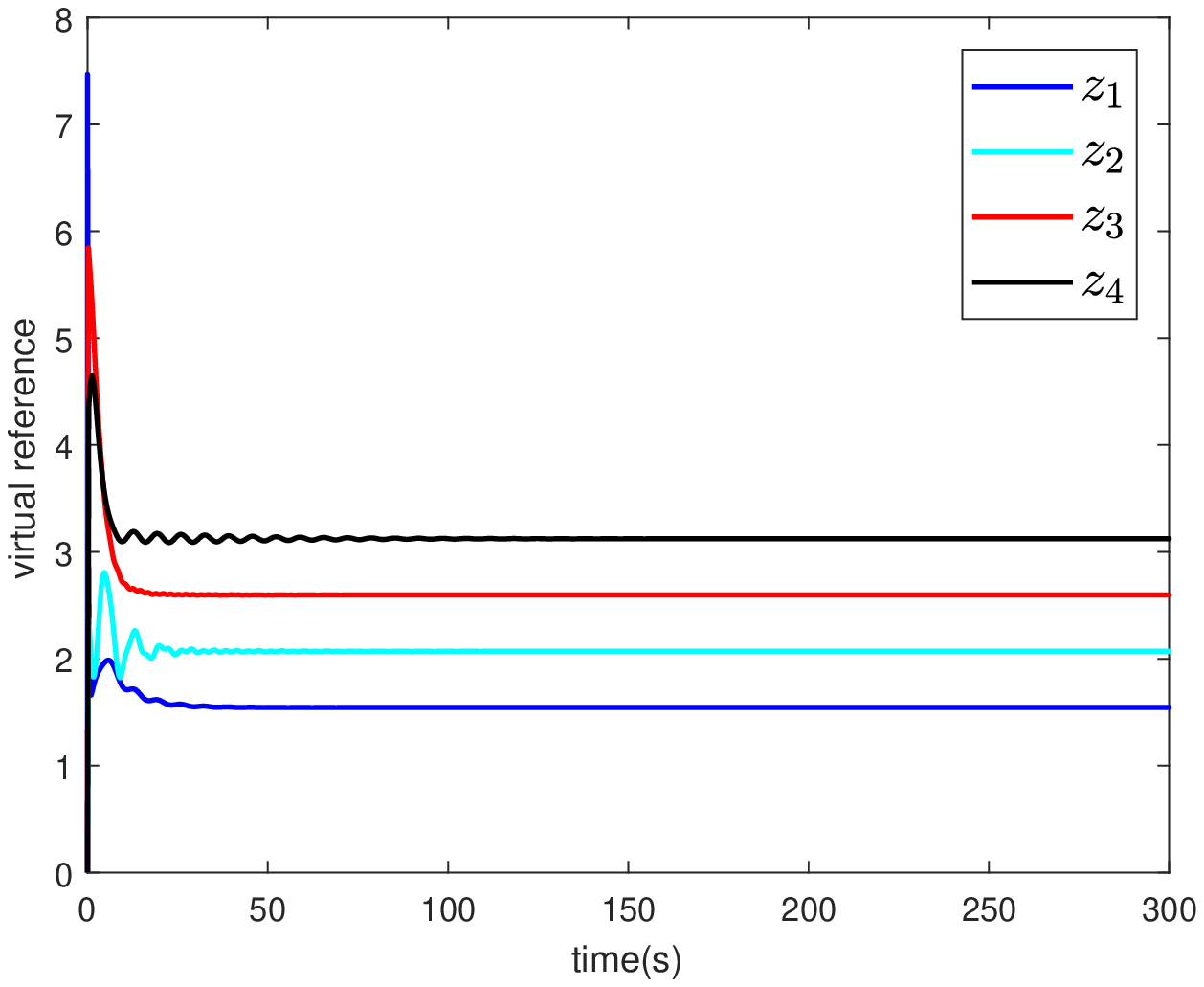}
	\caption{Profiles of $z_i(t)$ under the controller \eqref{ctr:online}.}\label{fig:simu-vanderpo-virtual}
\end{figure}

\begin{figure}
	\centering
	\includegraphics[width=0.4\textwidth]{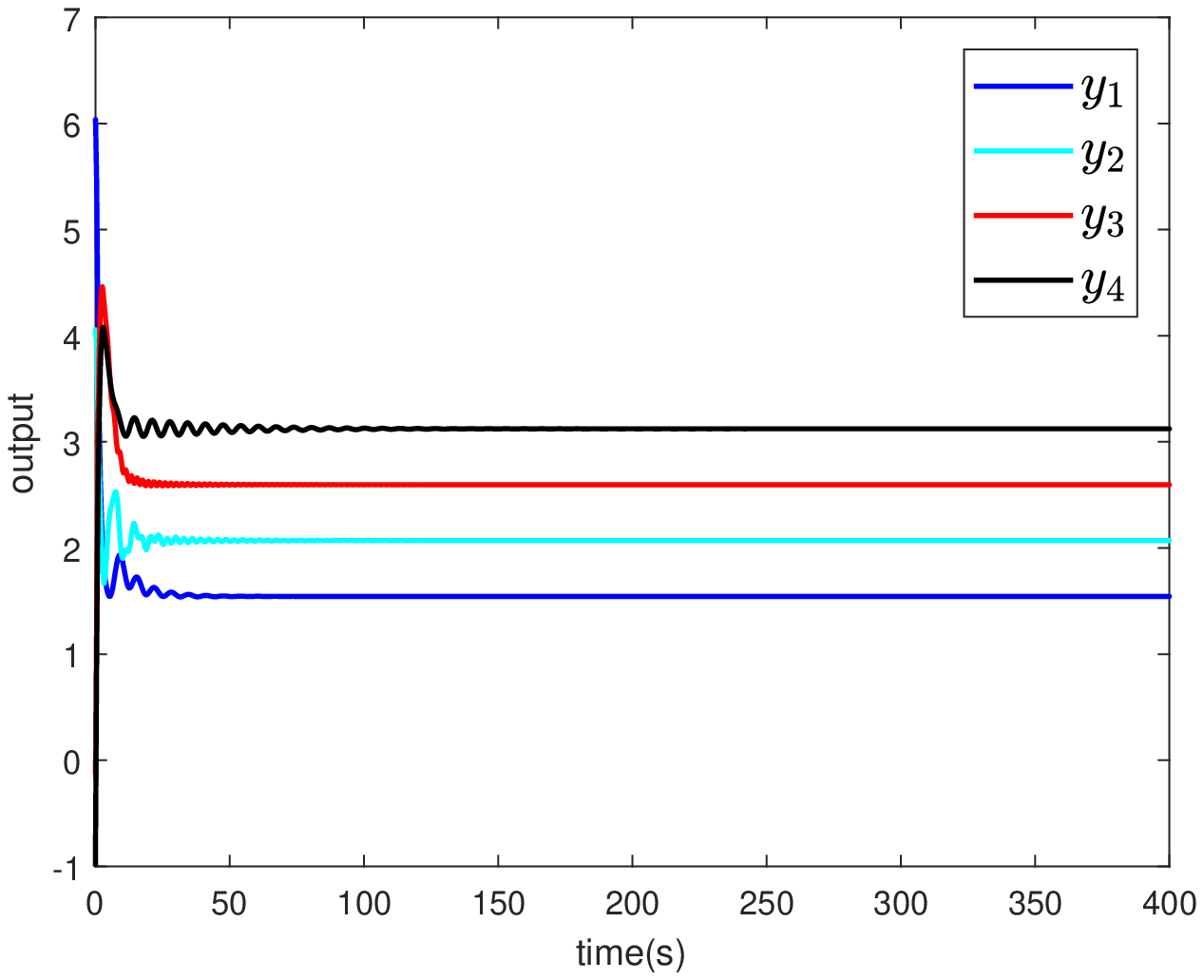}
	\caption{Profiles of $y_i(t)$ under the controller \eqref{ctr:online}.}\label{fig:simu-vanderpol-output}
\end{figure}

We let $p_i=0.1$, $q_i=1$, $y_{i0}=i$ and set the system parameters in agents as $a_i=b_i=1$, $\mu_1=0.1,\, \mu_2=-0.1$, $v_i(0)=\mbox{col}(0,\,2)$ for $i=1,\, \dots,\, 4$. Assumptions \ref{ass:convex+monotone}--\ref{ass:lip-extended} can be practically verified. Moreover,  the Nash equilibrium of this noncooperative game is $y^*=\mbox{col}(2.42,\,3.47,\,4.53,\,5.58)$ by numerical computations. According to Theorems \ref{thm:main} and \ref{thm:main-online}, the Nash equilibrium seeking problem for these agents can be solved by a distributed controller of the form \eqref{ctr:offline} or \eqref{ctr:online}.    

For simulations, we use the controller \eqref{ctr:online} using only real-time gradients.  Choose $\alpha=4$ for the virtual game dynamics \eqref{sys:generator} and $k_{1i}=-4 $, $k_{2i}=-4$, $\Lambda_{i}=5I_4$, $\e=0.8$ for the adaptive tracking controller. All initials are randomly chosen. 
Applying controller \eqref{ctr:offline} to agent \eqref{sys:agent}, the profiles of $z_i(t)$ and agent output $y_i(t)$ are shown in Figs.~\ref{fig:simu-vanderpo-virtual} and \ref{fig:simu-vanderpol-output}. It can be found that the Nash equilibrium $y^*$ is quickly reproduced even with real-time gradients, while the output $y_i(t)$ converges to the expected steady-state output strategy $y_i^*$ irrespective of the unknown nonlinearity $\Delta_i$ and external disturbance $d_i(t)$. 

\begin{figure}
	\centering
	\subfigure[]{
		\includegraphics[width=0.21\textwidth]{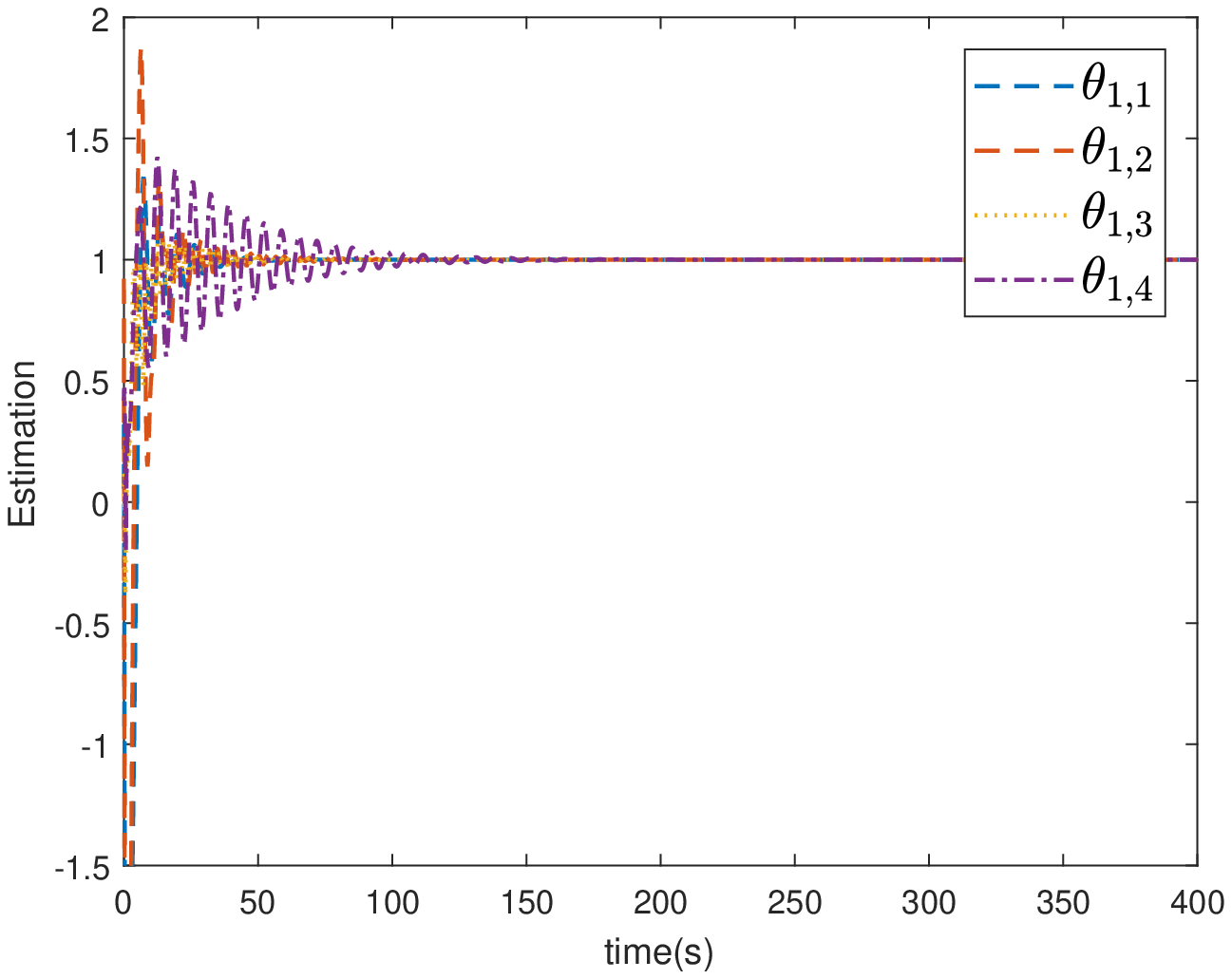}
	} 
	\subfigure[]{
		\includegraphics[width=0.21\textwidth]{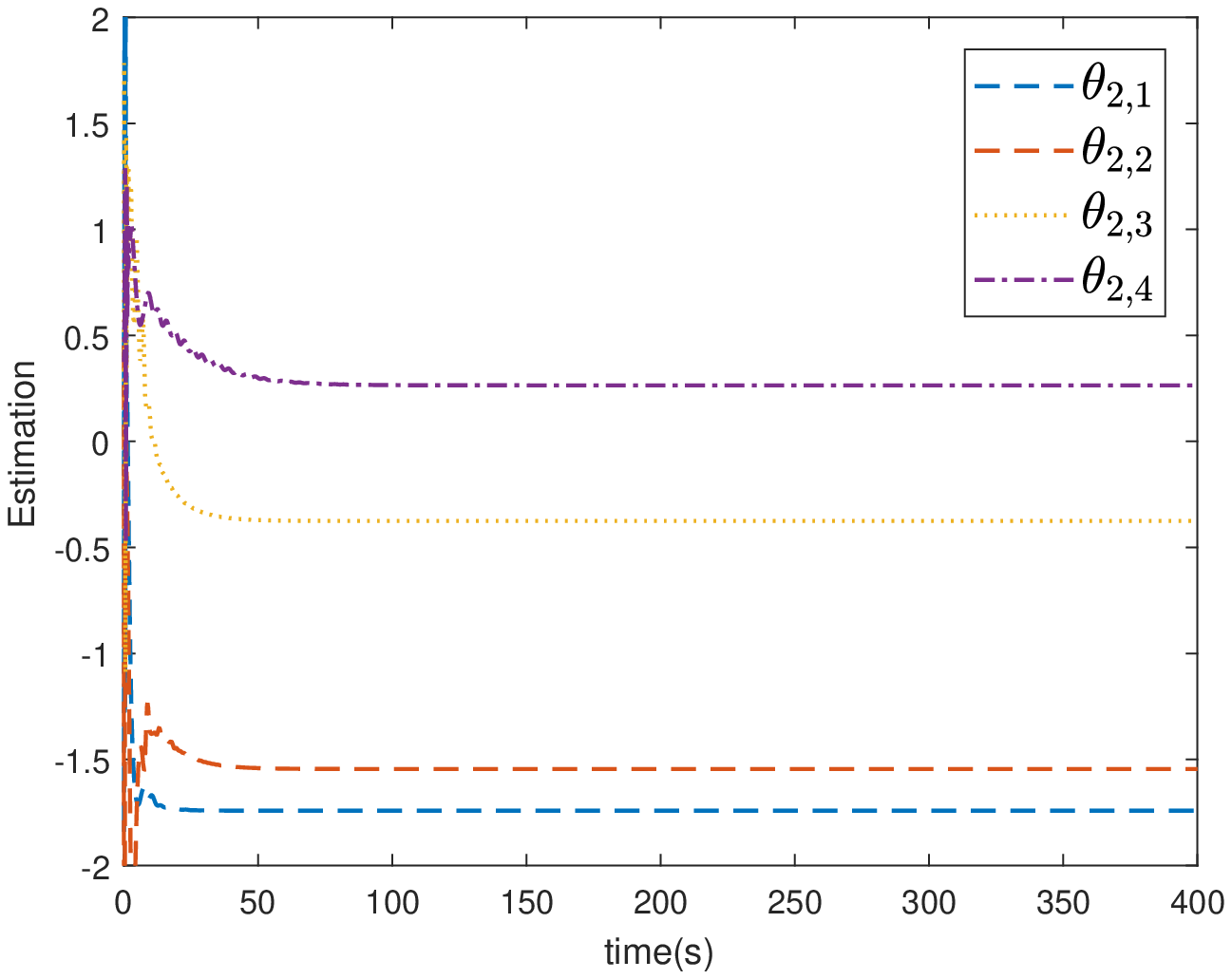}
	}\\
	\subfigure[]{
		\includegraphics[width=0.21\textwidth]{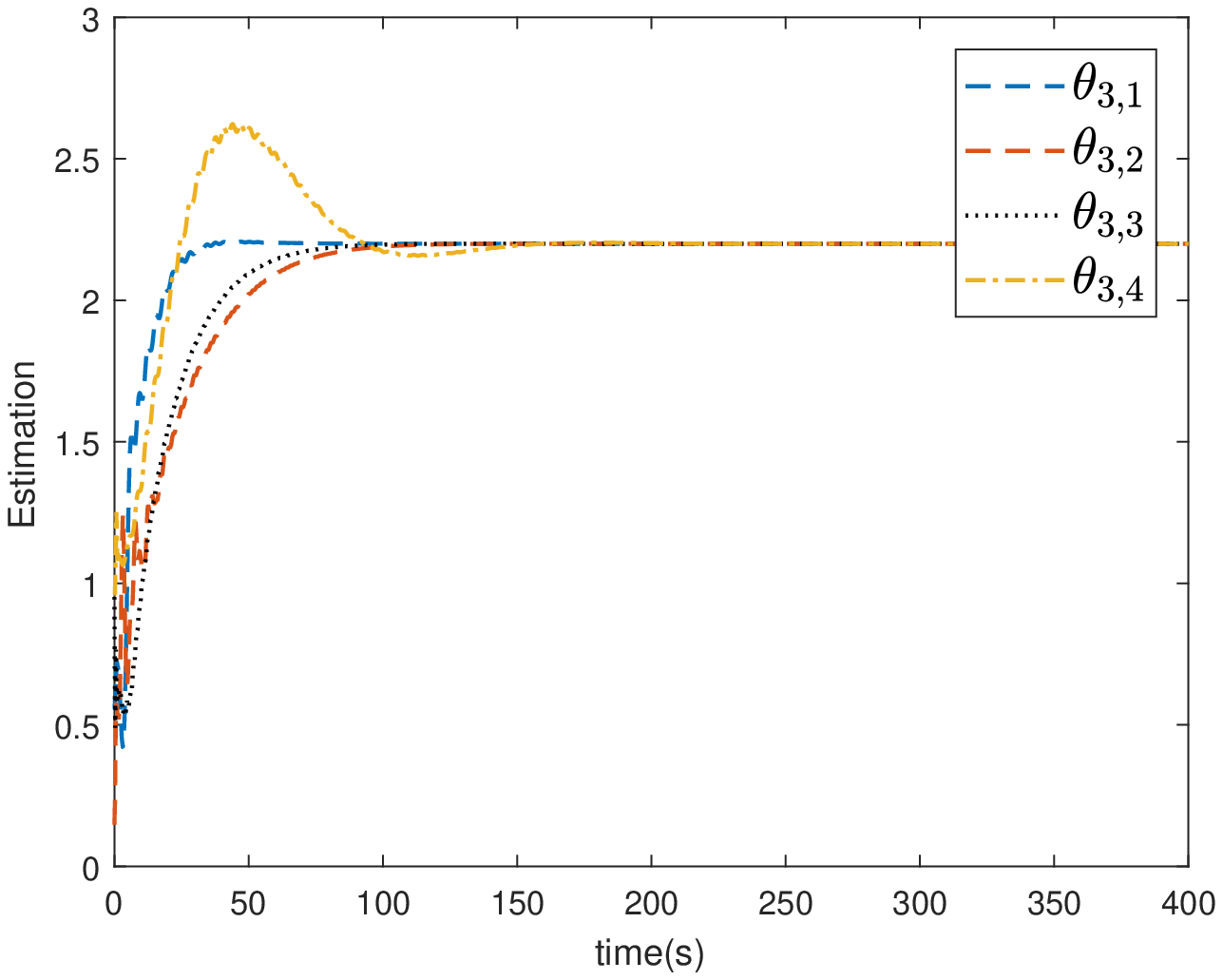}
	} 
	\subfigure[]{
		\includegraphics[width=0.21\textwidth]{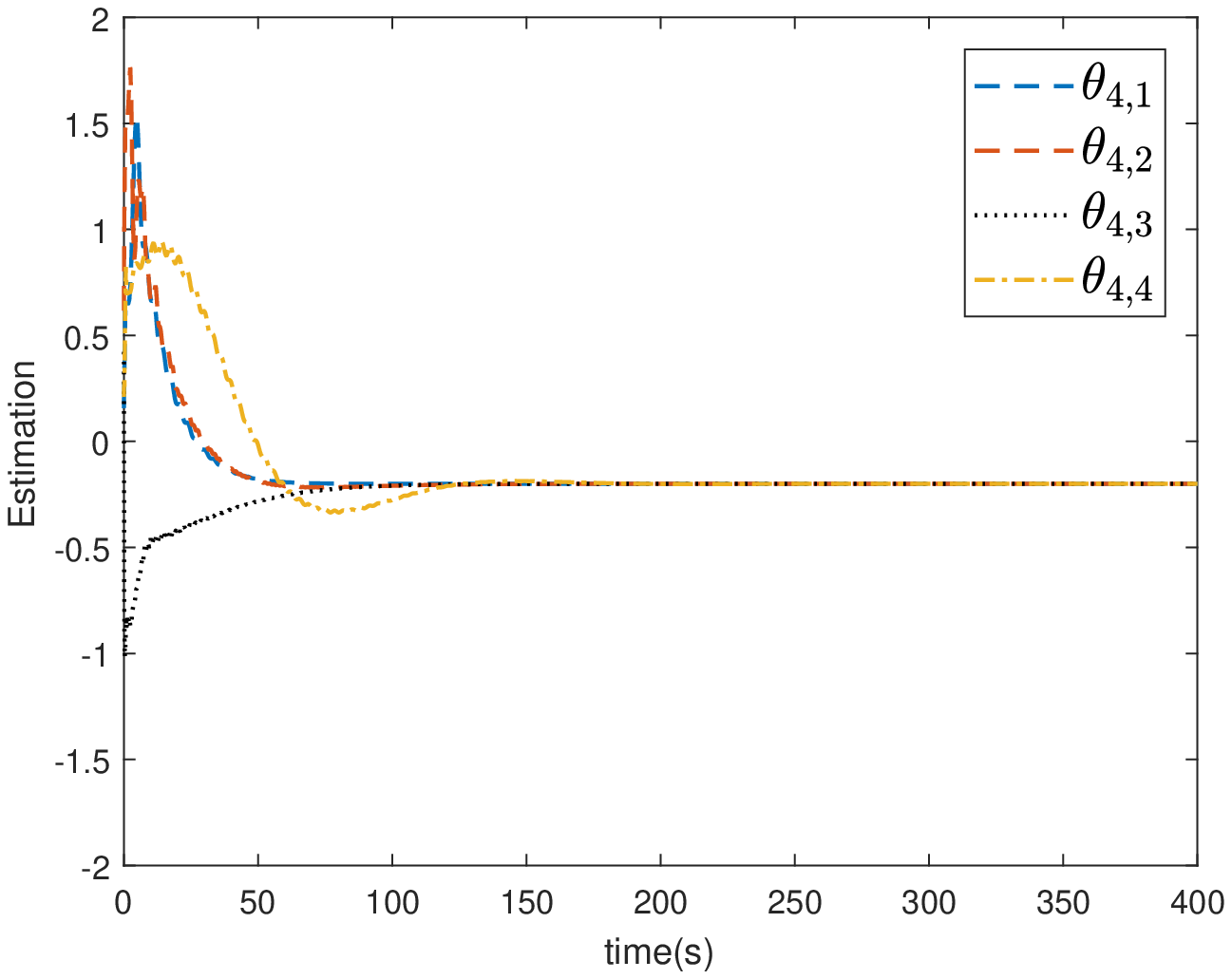}
	}
	\caption{Profiles of $\hat \theta_{i,j}(t)$ under the controller \eqref{ctr:online}.}\label{fig:simu-vanderpol-parameter}
\end{figure}

To explore the parameter convergence issue, we resort to Theorem \ref{thm:parameter} and Remark \ref{rem:parameter} and conclude that the estimators $\hat \theta_{1,i}$, $\hat \theta_{3,i}$, $\hat \theta_{4,i}$ will converge to their true values, while $\hat \theta_{2,i}$ may fail. These conclusions can be confirmed by Fig.~\ref{fig:simu-vanderpol-parameter}. 

%\begin{figure}
%	\centering
%	\includegraphics[width=0.45\textwidth]{game-adaptive-parametr-1.eps}
%	\caption{Profiles of $\hat \theta_{j,i}(t)$ with $j\neq 2$ under the controller \eqref{ctr:offline}.}\label{fig:parameter-1}
%\end{figure}
%
%
%\begin{figure}
%	\centering
%	\includegraphics[width=0.45\textwidth]{game-adaptive-parametr-2.eps}
%	\caption{Profiles of $\hat \theta_{2,i}(t)$ under the controller \eqref{ctr:offline}.}\label{fig:parameter-2}
%\end{figure}
%

\section{Conclusions}\label{sec:con}
A Nash equilibrium seeking problem has been considered for a typical class of high-order nonlinear systems with unknown dynamics. Following an embedded control procedure, we have developed a distributed adaptive controller to solve this problem under standard assumptions. The parameter convergence issue has also been addressed under some PE conditions. Output feedback controls and coupling constraints will be considered in our future work. 
	
\bibliographystyle{ieeetr}
\bibliography{ne_seeking_adaptive}

\begin{thebibliography}{10}

\bibitem{basar1999dynamic}
T.~Basar and G.~J. Olsder, {\em Dynamic Noncooperative Game Theory (2nd)}.
\newblock Philadelphia: SIAM, 1999.

\bibitem{stankovic2011distributed}
M.~S. Stankovic, K.~H. Johansson, and D.~M. Stipanovic, ``Distributed seeking
  of {Nash} equilibria with applications to mobile sensor networks,'' {\em IEEE
  Transactions on Automatic Control}, vol.~57, no.~4, pp.~904--919, 2011.

\bibitem{salehisadaghiani2016distributed}
F.~Salehisadaghiani and L.~Pavel, ``Distributed {Nash} equilibrium seeking: A
  gossip-based algorithm,'' {\em Automatica}, vol.~72, pp.~209--216, 2016.

\bibitem{koshal2016distributed}
J.~Koshal, A.~Nedi{\'c}, and U.~V. Shanbhag, ``Distributed algorithms for
  aggregative games on graphs,'' {\em Operations Research}, vol.~64, no.~3,
  pp.~680--704, 2016.

\bibitem{lou2016nash}
Y.~Lou, Y.~Hong, L.~Xie, G.~Shi, and K.~H. Johansson, ``Nash equilibrium
  computation in subnetwork zero-sum games with switching communications,''
  {\em IEEE Transactions on Automatic Control}, vol.~61, no.~10,
  pp.~2920--2935, 2016.

\bibitem{ye2017distributed}
M.~Ye and G.~Hu, ``Distributed {Nash} equilibrium seeking by a consensus based
  approach,'' {\em IEEE Transactions on Automatic Control}, vol.~62, no.~9,
  pp.~4811--4818, 2017.

\bibitem{zeng2019generalized}
X.~Zeng, J.~Chen, S.~Liang, and Y.~Hong, ``Generalized {Nash} equilibrium
  seeking strategy for distributed nonsmooth multi-cluster game,'' {\em
  Automatica}, vol.~103, pp.~20--26, 2019.

\bibitem{gadjov2019passivity}
D.~Gadjov and L.~Pavel, ``A passivity-based approach to {Nash} equilibrium
  seeking over networks,'' {\em IEEE Transactions on Automatic Control},
  vol.~64, no.~3, pp.~1077--1092, 2019.

\bibitem{de2019distributed}
C.~De~Persis and S.~Grammatico, ``Distributed averaging integral {Nash}
  equilibrium seeking on networks,'' {\em Automatica}, vol.~110, p.~108548,
  2019.

\bibitem{yi2019operator}
P.~Yi and L.~Pavel, ``An operator splitting approach for distributed
  generalized {Nash} equilibria computation,'' {\em Automatica}, vol.~102,
  pp.~111--121, 2019.

\bibitem{zhu2013coverage}
M.~Zhu and S.~Martínez, ``Distributed coverage games for energy-aware mobile
  sensor networks,'' {\em SIAM Journal on Control and Optimization}, vol.~51,
  no.~1, pp.~1--27, 2013.

\bibitem{frihauf2011nash}
P.~Frihauf, M.~Krstic, and T.~Basar, ``Nash equilibrium seeking in
  noncooperative games,'' {\em IEEE Transactions on Automatic Control},
  vol.~57, no.~5, pp.~1192--1207, 2011.

\bibitem{laraki2013higher}
R.~Laraki and P.~Mertikopoulos, ``Higher-order game dynamics,'' {\em Journal of
  Economic Theory}, vol.~148, no.~6, pp.~2666--2695, 2013.

\bibitem{fabiani2019nash}
F.~Fabiani and A.~Caiti, ``Nash equilibrium seeking in potential games with
  double-integrator agents,'' in {\em 2019 18th European Control Conference
  (ECC)}, pp.~548--553, IEEE, 2019.

\bibitem{ibrahim2019nash}
A.~R. Ibrahim and T.~Hayakawa, ``Nash equilibrium seeking with linear
  time-invariant dynamic agents,'' in {\em 2019 American Control Conference
  (ACC)}, pp.~1202--1207, IEEE, 2019.

\bibitem{romano2019dynamic}
A.~Romano and L.~Pavel, ``Dynamic {NE} seeking for multi-integrator networked
  agents with disturbance rejection,'' {\em IEEE Transactions on Control of
  Network Systems}, vol.~7, no.~1, pp.~129--139, 2019.

\bibitem{bianchi2019continuous}
M.~Bianchi and S.~Grammatico, ``Continuous-time fully distributed generalized
  {Nash} equilibrium seeking for multi-integrator agents,'' {\em arXiv preprint
  arXiv:1911.12266}, 2019.

\bibitem{de2019feedback}
C.~De~Persis and N.~Monshizadeh, ``A feedback control algorithm to steer
  networks to a {Cournot}--{Nash} equilibrium,'' {\em IEEE Transactions on
  Control of Network Systems}, vol.~6, no.~4, pp.~1486--1497, 2019.

\bibitem{deng2019distributed}
Z.~Deng and S.~Liang, ``Distributed algorithms for aggregative games of
  multiple heterogeneous {Euler}--{Lagrange} systems,'' {\em Automatica},
  vol.~99, pp.~246--252, 2019.

\bibitem{zhang2019distributed}
Y.~Zhang, S.~Liang, X.~Wang, and H.~Ji, ``Distributed {Nash} equilibrium
  seeking for aggregative games with nonlinear dynamics under external
  disturbances,'' {\em IEEE Transactions on Cybernetics}, pp.~1--10, 2019.

\bibitem{ruszczynski2006nonlinear}
A.~Ruszczynski, {\em Nonlinear Optimization}.
\newblock Princeton: Princeton University Press, 2006.

\bibitem{godsil2001algebraic}
C.~Godsil and G.~Royle, {\em Algebraic Graph Theory}.
\newblock New York: Springer, 2001.

\bibitem{khalil2002nonlinear}
H.~K. Khalil, {\em Nonlinear Systems (3rd ed.)}.
\newblock Upper Saddle River: Prentice Hall, 2002.

\bibitem{facchinei2003finite}
F.~Facchinei and J.-S. Pang, {\em Finite-dimensional variational inequalities
  and complementarity problems}.
\newblock New York: Springer Science \& Business Media, 2003.

\bibitem{ren2008distributed}
W.~Ren and R.~Beard, {\em Distributed Consensus in Multi-vehicle Cooperative
  Control: Theory and Applications}.
\newblock London: Springer, 2008.

\bibitem{tang2019cyb}
Y.~{Tang}, Z.~{Deng}, and Y.~{Hong}, ``Optimal output consensus of high-order
  multiagent systems with embedded technique,'' {\em IEEE Transactions on
  Cybernetics}, vol.~49, no.~5, pp.~1768--1779, 2019.

\bibitem{tang2020optimal}
Y.~{Tang} and X.~{Wang}, ``Optimal output consensus for nonlinear multi-agent
  systems with both static and dynamic uncertainties,'' {\em IEEE Transactions
  on Automatic Control, to appear}, 2021.

\bibitem{girard2009hierarchical}
A.~Girard and G.~J. Pappas, ``Hierarchical control system design using
  approximate simulation,'' {\em Automatica}, vol.~45, no.~2, pp.~566--571,
  2009.

\bibitem{tang2013hierarchical}
Y.~Tang and Y.~Hong, ``Hierarchical distributed control design for multi-agent
  systems using approximate simulation,'' {\em Acta Automatica Sinica},
  vol.~39, no.~6, pp.~868--874, 2013.

\bibitem{krstic1995nonlinear}
M.~Krstic, I.~Kanellakopoulos, and P.~V. Kokotovic, {\em Nonlinear and Adaptive
  Control Design}.
\newblock New York: Wiley, 1995.

\bibitem{tang2020distributed}
Y.~{Tang}, ``Distributed optimal steady-state regulation for high-order
  multiagent systems with external disturbances,'' {\em IEEE Transactions on
  Systems, Man, and Cybernetics: Systems}, vol.~50, no.~11, pp.~4828--4835,
  2020.

\bibitem{ioannou1995robust}
P.~A. Ioannou and J.~Sun, {\em Robust Adaptive Control}.
\newblock Upper Saddle River: Prentice-Hall, 1995.

\bibitem{mazenc2009uniform}
F.~Mazenc, M.~De~Queiroz, and M.~Malisoff, ``Uniform global asymptotic
  stability of a class of adaptively controlled nonlinear systems,'' {\em IEEE
  Transactions on Automatic Control}, vol.~54, no.~5, pp.~1152--1158, 2009.

\bibitem{hu2014adaptive}
J.~Hu and W.~Zheng, ``Adaptive tracking control of leader--follower systems
  with unknown dynamics and partial measurements,'' {\em Automatica}, vol.~50,
  no.~5, pp.~1416--1423, 2014.

\bibitem{chen2015stabilization}
Z.~Chen and J.~Huang, {\em Stabilization and Regulation of Nonlinear Systems: A
  Robust and Adaptive Approach}.
\newblock Cham, Switzerland: Springer, 2015.

\bibitem{ortega1993asymptotic}
R.~Ortega and A.~Fradkov, ``Asymptotic stability of a class of adaptive
  systems,'' {\em International Journal of Adaptive Control and Signal
  Processing}, vol.~7, no.~4, pp.~255--260, 1993.

\bibitem{raafat1991survey}
F.~Raafat, ``Survey of literature on continuously deteriorating inventory
  models,'' {\em Journal of the Operational Research society}, vol.~42, no.~1,
  pp.~27--37, 1991.

\end{thebibliography}
	
\end{document}